\documentclass[twocolumn,
               aps,
               prd,   
               preprintnumbers,
               superscriptaddress,
               numbers,
               floatfix,
               sort&compress,
               nofootinbib,
               showpacs,
               colorlinks,
               linkcolor=blue,   
               citecolor=blue]{revtex4-2}

\usepackage{graphicx}
\usepackage{dcolumn}
\usepackage{bm}
\usepackage{hyperref}
\usepackage{verbatim}

\usepackage{graphicx}
\usepackage{dcolumn}
\usepackage{bm}
\usepackage{hyperref}
\usepackage{float}

\usepackage{amsmath}
\usepackage[caption=false]{subfig}
\usepackage{siunitx}
\usepackage{soul}

\def\<{\langle}
\def\>{\rangle}

\def\+{\dagger}

\def\U1A{U(1)$_{\rm A}$}

\def\ra{\rangle}
\def\la{\langle}

\newcommand{\be}{\begin{eqnarray}}
\newcommand{\ee}{\end{eqnarray}}
\newcommand{\beq}{\begin{equation}}
\newcommand{\eeq}{\end{equation}}
\newcommand{\exclude}[1]{}
\newcommand{\jcap}{JCAP}

\newcommand{\physrep}{Phys. Rep.}

\begin{document}


\title{Daily modulations and broadband strategy in axion searches: An application with CAST-CAPP detector}

\author{    F.~Caspers  }
\thanks{deceased}
\affiliation{European Organization for Nuclear Research (CERN), CH-1211 Gen\`eve, Switzerland}
\affiliation{European Scientific Institute (ESI), 74160 Archamps, France}

\author{    C.~M.~Adair}
\affiliation{Department of Physics and Astronomy, University of British Columbia, Vancouver V6T 1Z1, Canada}

\author{    K.~Altenm\"{u}ller}
\affiliation{Centro de Astropart\'{\i}culas y F\'{\i}sica de Altas Energ\'{\i}as (CAPA), Universidad de Zaragoza, 50009 Zaragoza, Spain}

\author{    V.~Anastassopoulos}
\affiliation{Physics Department, University of Patras, 26504 Patras, Greece}

\author{ S.~Arguedas Cuendis}
\affiliation{European Organization for Nuclear Research (CERN), CH-1211 Gen\`eve, Switzerland}

\author{    J.~Baier} 
\affiliation{Physikalisches Institut, Albert-Ludwigs-Universit\"{a}t Freiburg, 79104 Freiburg, Germany}

\author{    K.~Barth}
\affiliation{European Organization for Nuclear Research (CERN), CH-1211 Gen\`eve, Switzerland}

\author{    A.~Belov  }
\affiliation{Author affiliated with an institute covered by a cooperation agreement with CERN}

\author{  D.~Bozicevic }
\affiliation{University of Rijeka, Faculty of Engineering, 51000 Rijeka, Croatia}

\author{    H.~Br\"auninger  } 
\thanks{deceased}
\affiliation{Max-Planck-Institut f\"{u}r Extraterrestrische Physik, D-85741 Garching, Germany}

\author{    G.~Cantatore  } 
\affiliation{Istituto Nazionale di Fisica Nucleare (INFN), Sezione di Trieste, 34127 Trieste, Italy}
\affiliation{Universit\`a di Trieste, 34127 Trieste, Italy}

\author{    J.~F.~Castel  } 
\affiliation{Centro de Astropart\'{\i}culas y F\'{\i}sica de Altas Energ\'{\i}as (CAPA), Universidad de Zaragoza, 50009 Zaragoza, Spain}

\author{    S.~A.~\c{C}etin  }
\affiliation{Istinye University, Department of Basic Sciences, 34396 Sariyer, Istanbul, T\"{u}rkiye}

\author{     W.~Chung  }
\affiliation{Center for Axion and Precision Physics Research, Institute for Basic Science (IBS), Daejeon 34141, Republic of Korea}

\author{     H.~Choi  }
\affiliation{Department of Physics, Korea Advanced Institute of Science and Technology (KAIST), Daejeon 34141, Republic of Korea}

\author{     J.~Choi  }
\affiliation{Center for Axion and Precision Physics Research, Institute for Basic Science (IBS), Daejeon 34141, Republic of Korea}

\author{    T.~Dafni  }
\affiliation{Centro de Astropart\'{\i}culas y F\'{\i}sica de Altas Energ\'{\i}as (CAPA), Universidad de Zaragoza, 50009 Zaragoza, Spain}

\author{    M.~Davenport  }
\affiliation{European Organization for Nuclear Research (CERN), CH-1211 Gen\`eve, Switzerland}

\author{    A.~Dermenev  }
\affiliation{Author affiliated with an institute covered by a cooperation agreement with CERN}

\author{	K.~Desch}
\affiliation{Physikalisches Institut, University of Bonn, 53115 Bonn, Germany}

\author{    B.~D\"obrich  }
\affiliation{Max-Planck-Institut f\"{u}r Physik (Werner-Heisenberg-Institut), Boltzmannstr. 8, 85748 Garching bei M\"{u}nchen, Germany}

\author{    H.~Fischer  } 
\affiliation{Physikalisches Institut, Albert-Ludwigs-Universit\"{a}t Freiburg, 79104 Freiburg, Germany}

\author{    W.~Funk  } 
\affiliation{European Organization for Nuclear Research (CERN), CH-1211 Gen\`eve, Switzerland}

\author{    J.~Galan}
\affiliation{Centro de Astropart\'{\i}culas y F\'{\i}sica de Altas Energ\'{\i}as (CAPA), Universidad de Zaragoza, 50009 Zaragoza, Spain}

\author{    A.~Gardikiotis  }
\affiliation{Istituto Nazionale di Fisica Nucleare (INFN), Sezione di Padova, 35131 Padova, Italy}

\author{    S.~Gninenko  }  
\affiliation{Author affiliated with an institute covered by a cooperation agreement with CERN}

\author{    J.~Golm  }
\affiliation{European Organization for Nuclear Research (CERN), CH-1211 Gen\`eve, Switzerland}
\affiliation{Institute for Optics and Quantum Electronics, Friedrich Schiller University Jena, 07743 Jena, Germany}

\author{    M.~D.~Hasinoff  }
\affiliation{Department of Physics and Astronomy, University of British Columbia, V6T 1Z1 Vancouver, Canada}

\author{    D.~H.~H.~Hoffmann  }
\affiliation{Xi'An Jiaotong University, School of Science, Xi'An, 710049, China}

\author{   D.~D\'{\i}ez Ib\'a\~{n}ez}
\affiliation{Centro de Astropart\'{\i}culas y F\'{\i}sica de Altas Energ\'{\i}as (CAPA), Universidad de Zaragoza, 50009 Zaragoza, Spain}

\author{   I.~G.~Irastorza  } 
\affiliation{Centro de Astropart\'{\i}culas y F\'{\i}sica de Altas Energ\'{\i}as (CAPA), Universidad de Zaragoza, 50009 Zaragoza, Spain}

\author{   K.~Jakov\v ci\' c  } 
\affiliation{Rudjer Bo\v{s}kovi\'{c} Institute, 10000 Zagreb, Croatia}

\author{	J.~Kaminski} 
\affiliation{Physikalisches Institut, University of Bonn, 53115 Bonn, Germany}

\author{   M.~Karuza } 
\affiliation{Istituto Nazionale di Fisica Nucleare (INFN), Sezione di Trieste, 34127 Trieste, Italy}
\affiliation{University of Rijeka, Faculty of Physics, 51000 Rijeka, Croatia}
\affiliation{University of Rijeka, Photonics and Quantum Optics Unit, Center of Excellence for Advanced Materials and Sensing Devices, and Centre for Micro and Nano Sciences and Technologies, 51000 Rijeka, Croatia}

\author{ 	C. Krieger} 
\affiliation{Institute of Experimental Physics, University of Hamburg, 22761 Hamburg, Germany}

\author{     \c{C}.~Kutlu  }
\affiliation{Center for Axion and Precision Physics Research, Institute for Basic Science (IBS), Daejeon 34141, Republic of Korea}
\affiliation{Department of Physics, Korea Advanced Institute of Science and Technology (KAIST), Daejeon 34141, Republic of Korea}

\author{  B.~Laki\'{c}}
\thanks{deceased}
\affiliation{Rudjer Bo\v{s}kovi\'{c} Institute, 10000 Zagreb, Croatia}

\author{    J.~M.~Laurent  } 
\affiliation{European Organization for Nuclear Research (CERN), CH-1211 Gen\`eve, Switzerland}

\author{    J.~Lee  }
\affiliation{Department of Physics, Korea Advanced Institute of Science and Technology (KAIST), Daejeon 34141, Republic of Korea}

\author{   S.~Lee  }
\affiliation{Center for Axion and Precision Physics Research, Institute for Basic Science (IBS), Daejeon 34141, Republic of Korea}

\author{    G.~Luz\'on  }
\affiliation{Centro de Astropart\'{\i}culas y F\'{\i}sica de Altas Energ\'{\i}as (CAPA), Universidad de Zaragoza, 50009 Zaragoza, Spain}


\author{    C.~Margalejo  }
\affiliation{Centro de Astropart\'{\i}culas y F\'{\i}sica de Altas Energ\'{\i}as (CAPA), Universidad de Zaragoza, 50009 Zaragoza, Spain}

\author{    M.~Maroudas }
\email{marios.maroudas@cern.ch}
\affiliation{Institute of Experimental Physics, University of Hamburg, 22761 Hamburg, Germany}

\author{    L.~Miceli  }
\affiliation{Center for Axion and Precision Physics Research, Institute for Basic Science (IBS), Daejeon 34141, Republic of Korea}

\author{    H.~Mirallas  } 
\affiliation{Centro de Astropart\'{\i}culas y F\'{\i}sica de Altas Energ\'{\i}as (CAPA), Universidad de Zaragoza, 50009 Zaragoza, Spain}

\author{    L.~Obis}
\affiliation{Centro de Astropart\'{\i}culas y F\'{\i}sica de Altas Energ\'{\i}as (CAPA), Universidad de Zaragoza, 50009 Zaragoza, Spain}

\author{A.~\"{O}zbey}
\affiliation{Department of Mechanical Engineering, Istanbul University-Cerrahpaşa, Istanbul, T\"{u}rkiye}

\author{    K.~\"{O}zbozduman  }
\email{kaan.ozbozduman@cern.ch}
\affiliation{Bogazici University, Physics Department, 34342 Bebek, Istanbul, T\"{u}rkiye}

\author{    M.~J.~Pivovaroff  }
\affiliation{Lawrence Livermore National Laboratory, Livermore, California 94550, USA}

\author{    M.~Rosu  }
\affiliation{Extreme Light Infrastructure - Nuclear Physics (ELI-NP), 077125 Magurele, Romania}

\author{    J.~Ruz  } 
\affiliation{Lawrence Livermore National Laboratory, Livermore, CA 94550, USA}

\author{    E.~Ruiz-Ch\'oliz  }
\affiliation{Institut f\"{u}r Physik, Johannes Gutenberg Universit\"{a}t Mainz, 55128 Mainz, Germany}

\author{    S.~Schmidt  } 
\affiliation{Physikalisches Institut, University of Bonn, 53115 Bonn, Germany}


\author{    Y.~K.~Semertzidis  }
\affiliation{Center for Axion and Precision Physics Research, Institute for Basic Science (IBS), Daejeon 34141, Republic of Korea}
\affiliation{Department of Physics, Korea Advanced Institute of Science and Technology (KAIST), Daejeon 34141, Republic of Korea}

\author{    S.~K.~Solanki  }
\affiliation{Max-Planck-Institut f\"{u}r Sonnensystemforschung, 37077 G\"{o}ttingen, Germany}

\author{    L.~Stewart  }
\affiliation{European Organization for Nuclear Research (CERN), CH-1211 Gen\`eve, Switzerland}

\author{    I.~Tsagris} 
\affiliation{Physics Department, University of Patras, 26504 Patras, Greece}

\author{    T.~Vafeiadis  }
\affiliation{European Organization for Nuclear Research (CERN), CH-1211 Gen\`eve, Switzerland}

\author{    J.~K.~Vogel  }
\affiliation{Centro de Astropart\'{\i}culas y F\'{\i}sica de Altas Energ\'{\i}as (CAPA), Universidad de Zaragoza, 50009 Zaragoza, Spain}
\affiliation{Lawrence Livermore National Laboratory, Livermore, CA 94550, USA}
\affiliation{Present address: Fakult\"{a}t f\"{u}r Physik, TU Dortmund, Otto-Hahn-Str. 4, Dortmund D-44221, Germany}

\author{    M.~Vretenar  }
\affiliation{University of Rijeka, Faculty of Physics, 51000 Rijeka, Croatia}

\author{    S.~Youn  }
\affiliation{Center for Axion and Precision Physics Research, Institute for Basic Science (IBS), Daejeon 34141, Republic of Korea}

\author{    A.~Zhitnitsky  }
\email{arz@phas.ubc.ca}
\affiliation{Department of Physics and Astronomy, University of British Columbia, V6T 1Z1 Vancouver, Canada}

\author{    K.~Zioutas  }
\affiliation{Physics Department, University of Patras, 26504 Patras, Greece}

\date{\today}

\begin{abstract}
It has been previously advocated that the presence of the daily and annual modulations of the axion flux on the Earth's surface may dramatically change the strategy of the axion searches. The arguments were based on the so-called Axion Quark Nugget (AQN) dark matter model which was originally put forward to explain the similarity of the dark and visible cosmological matter densities $\Omega_{\rm dark}\sim \Omega_{\rm visible}$. In this framework, the population of galactic axions with mass $ 10^{-6} {\rm eV}\lesssim m_a\lesssim 10^{-3}{\rm eV}$ and velocity $\langle v_a\rangle\sim 10^{-3} c$ will be accompanied by axions with typical velocities $\langle v_a\rangle\sim 0.6 c$ emitted by AQNs. Furthermore, in this framework, it has also been argued that the AQN-induced axion daily modulation (in contrast with the conventional WIMP paradigm) could be as large as $(10-20)\%$, representing the main motivation for the present investigation. We argue that the daily modulations along with the broadband detection strategy can be very useful tools for the discovery of such relativistic axions. The data from the CAST-CAPP detector have been used following such arguments. Unfortunately, due to the dependence of the amplifier chain on temperature-dependent gain drifts and other factors, we could not conclusively show the presence or absence of a dark sector-originated daily modulation. However, this proof of principle analysis procedure can serve as a reference for future studies.
\end{abstract}

\keywords{Axion Quark Nuggets, daily modulation, haloscope, CAST-CAPP}

\maketitle

\section{Introduction}
\label{sect:introduction}

The Peccei-Quinn mechanism, accompanied by axions, offers the most compelling solution to the strong $\cal{CP}$ problem; see foundational works \cite{1977PhRvD..16.1791P, 1978PhRvL..40..223W, 1978PhRvL..40..279W, KSVZ1, KSVZ2, DFSZ1, DFSZ2} and recent reviews \cite{vanBibber:2006rb, Asztalos:2006kz, Sikivie:2008, Raffelt:2006cw, Sikivie:2009fv, Rosenberg:2015kxa, Marsh:2015xka, Graham:2015ouw, Battesti:2018bgc, Irastorza:2018dyq, Sikivie:2020zpn}. The conventional approach for producing dark matter (DM) axions relies on either the misalignment mechanism—where the cosmological field $\theta(t)$ oscillates and emits cold axions before reaching a minimum or the decay of topological defects, as outlined in recent reviews \cite{vanBibber:2006rb, Asztalos:2006kz, Sikivie:2008, Raffelt:2006cw, Sikivie:2009fv, Rosenberg:2015kxa, Marsh:2015xka, Graham:2015ouw, Battesti:2018bgc, Irastorza:2018dyq, Sikivie:2020zpn}. In both cases, the resulting axions are nonrelativistic, with velocities of approximately $\langle v_a \rangle \sim 10^{-3} c$.

Beyond these established mechanisms, recent studies \cite{Fischer:2018niu, Liang:2018ecs, Lawson:2019cvy, Liang:2019lya, Budker:2019zka} have proposed a novel mechanism for axion production based on the axion quark nugget (AQN) DM model \cite{Zhitnitsky:2002qa}, briefly reviewed in Sec.\ref{AQN}. This model shares similarities with Witten’s quark-nugget model \cite{Witten:1984rs}, but is ``cosmologically dark" due to its low cross section-to-mass ratio, which reduces observable consequences of what is otherwise a strongly interacting DM candidate. An essential feature of the AQN model is the production of relativistic axions with velocities of $\langle v_a \rangle \sim 0.6 c$, in contrast to conventional mechanisms where axion velocities are typical of galactic DM ($\langle v_a \rangle \sim 10^{-3} c$), as detailed in Sec.\ref{AQN-induced}.

This unique characteristic necessitates a broadband search for axion detection, markedly changing the detection strategy. Such broadband searches could incorporate additional elements, including a specific analysis of daily modulation effects (detailed in Sec.~\ref{strategy}).

Relativistic axion production has also been discussed outside of the AQN framework, as seen in Ref. \cite{Dror:2021nyr}, where several production mechanisms were suggested and collectively termed the cosmic axion background (CaB). Thus, our proposed broadband search strategy could apply more broadly and is not limited solely to the AQN model.

Despite these parallels, several distinctions exist between AQN-induced axions and CaB axions as considered in Ref. \cite{Dror:2021nyr}. AQN-induced axions arise from interactions between AQNs and the Earth’s material, meaning they are largely localized near the Earth's (or other celestial bodies') surface. In contrast, the CaB axions in Ref. \cite{Dror:2021nyr} are cosmologically distributed throughout the Universe. Consequently, the dominant portion of the relativistic axions (which is the topic of the present work) should be related to the AQN-induced events with a well-defined estimate for the axion density on the Earth's surface as reviewed in Sec.~\ref{AQN-induced}.

Another key difference concerns daily modulation. The CaB-induced daily modulation observed in Ref.~\cite{Dror:2021nyr} requires directional sensitivity in detectors, as the modulation arises from the shifting orientation of Earth's axis relative to the galactic wind. Although current instruments were not specifically designed for directional detection, recent studies like \cite{ADMX:2023rsk} have utilized Axion Dark Matter Experiment (ADMX) data to examine daily modulation due to CaB. For AQN-induced axions, an additional source of daily modulation arises from the changing mass (and size) of AQNs entering or exiting Earth's atmosphere, as discussed in Sec.~\ref{AQN-induced}. This additional modulation allows the study of daily modulation effects even with spherically symmetric, nondirectionally sensitive instruments.

Furthermore, the AQN-induced axion density near the Earth’s surface is expected to surpass the CaB axion density, which is constrained by the cosmic microwave background (CMB) photon density (see Sec.~\ref{AQN-induced} for estimates). Therefore, time modulation effects, which are the topic of the present study, will be dominated by AQN-induced axions.


One additional feature of AQN-induced axions that differs from CaB mechanisms in \cite{Dror:2021nyr} is that AQNs, being massive objects, produce a highly nonuniform flux on Earth. This nonuniformity, which is in contrast with the conventional DM models, and the CaB flux, can create "local flashes" of axions. This can temporarily increase their count by several orders of magnitude when an AQN impacts the Earth's surface within about \SI{100}{\km} of the detector (see Sec.~\ref{AQN} for further estimates).

This paper is organized as follows: Section \ref{AQN} provides an overview of the AQN framework, while Secs.\ref{AQN-induced} and \ref{strategy} describe the main properties of AQN-induced axions relevant to this study, including their spectrum and intensity at Earth's surface. Section \ref{signals} presents our estimates of daily modulation effects and suggests methods to distinguish genuine signals from noise. In Sec.\ref{sec:cast_capp_analysis}, we apply these analytical features to explore daily modulation in previously recorded CAST-CAPP data. We conclude in Sec.~\ref{conclusion}, with recommendations for future studies.

\section{AQN dark matter model: The basics}
\label{AQN}

The AQN model shares several key features with Witten’s quark nugget construction \cite{Witten:1984rs, Farhi:1984qu, DeRujula:1984axn}. Like quark nuggets, AQNs are ``cosmologically dark'' due to the very low cross-section-to-mass ratio relevant for cosmological interactions. This small ratio effectively limits the observable impacts of these otherwise strongly interacting DM candidates.

The AQN model, however, incorporates additional elements beyond the original quark nugget framework \cite{Witten:1984rs, Farhi:1984qu, DeRujula:1984axn}. First, AQNs are stabilized by axion domain walls, which form abundantly during the QCD phase transition and mitigate several issues associated with Witten's original model. Second, during this QCD transition, AQNs can form as either {\it matter} or {\it antimatter} nuggets, adding further diversity to their composition.


\begin{table*}
\captionsetup{justification=raggedright}
	\begin{tabular}{cccrcc} 
		\hline\hline
		  Property  && \begin{tabular} {@{}c@{}}{ Typical value or feature}~~~~~\end{tabular} \\\hline
		  AQN's mass~  $[M_N]$ &&         $M_N\approx 16\,g\,(B/10^{25})$     \cite{Zhitnitsky:2021iwg}     \\
		   Baryon charge constraints~   $ [B]  $   &&        $ B \geq 3\cdot 10^{24}  $     \cite{Zhitnitsky:2021iwg}    \\
		   Annihilation cross section~  $[\sigma]$ &&     $\sigma\approx\kappa\pi R^2\simeq 1.5\cdot 10^{-9} {\rm cm^2} \cdot  \kappa (R/2.2\cdot 10^{-5}\rm cm)^2$  ~~~~     \\
		  Density of AQNs~ $[n_{\rm AQN}]$         &&          $n_{\rm AQN} \sim 0.3\cdot 10^{-25} {\rm cm^{-3}} (10^{25}/B) $   \cite{Zhitnitsky:2021iwg} \\
		  Survival pattern during BBN &&       $\Delta B/B\ll 1$  \cite{Zhitnitsky:2006vt,Flambaum:2018ohm,SinghSidhu:2020cxw,Santillan:2020lbj} \\
		  Survival pattern during CMB &&           $\Delta B/B\ll 1$ \cite{Zhitnitsky:2006vt,Lawson:2018qkc,SinghSidhu:2020cxw} \\
		  Survival pattern during post-recombination &&   $\Delta B/B\ll 1$ \cite{Ge:2019voa} \\\hline
	\end{tabular}
	\caption{Basic  properties of the AQNs} 
	\label{tab:basics}
\end{table*}
The presence of antimatter nuggets in the AQN  framework is an inevitable and direct consequence of the $\cal{CP}$ violating axion field which is present in the system during the  QCD time. As a result of this feature, the DM density $\Omega_{\rm DM}$, and the visible density $\Omega_{\rm visible}$, will automatically assume the same order of magnitude densities  $\Omega_{\rm DM}\sim \Omega_{\rm visible}$  irrespectively to the parameters of the model, such as the axion mass $m_a$. This feature represents a generic property of the construction \cite{Zhitnitsky:2002qa} as both components, the visible, and the dark are proportional to the same fundamental dimensional constant of the theory, the $\Lambda_{\rm QCD}$. 
  
For detailed discussions on the formation of AQNs, the generation of baryon asymmetry, and the survival mechanisms of these nuggets in the hostile environment of the early Universe, we refer to the original studies \cite{Liang:2016tqc, Ge:2017ttc, Ge:2017idw, Ge:2019voa}. Additionally, a recent review article \cite{Zhitnitsky:2021iwg} provides insights into the formation and survival pattern of AQNs during the early stages of the evolution, including the CMB Big Bang Nucleosynthesis (BBN), and recombination epochs.

In this work, we take an agnostic stance and assume that {\it antimatter}-based nuggets exist in our Universe today and have existed since earlier cosmic times, independent of any specific formation mechanism. This assumption aligns with all current cosmological, astrophysical, and terrestrial constraints, provided that the average baryon charge of these nuggets is sufficiently large, as we review below.
 
The most stringent direct detection limit\footnote{Nondetection of etching tracks in ancient mica offers an indirect constraint on the flux of DM nuggets with masses above $M > 55$g \cite{Jacobs:2014yca}. This constraint assumes that all nuggets have identical mass, which does not apply to the AQN model.} comes from the IceCube Observatory; see Appendix A in Ref.~\cite{Lawson:2019cvy}:
\be
\label{direct}
\la B \ra > 3\cdot 10^{24} ~~~[{\rm direct ~ (non)detection ~constraint]}.
\ee

Similar limits can also be obtained from ANITA and geothermal constraints, which are consistent with Eq.~\ref{direct}, as estimated in Ref.~\cite{Gorham:2012hy}.

\exclude{
The authors of Ref. \cite{SinghSidhu:2020cxw} considered a generic constraint for the nuggets made of antimatter (ignoring all essential specifics of the AQN model such as quark matter  CS phase of the nugget's core). Our constraints (Eq.~\ref{direct}) are consistent with their findings including the CMB and BBN, and others, except the constraints derived from the so-called ``Human Detectors". As explained in \cite{Ge:2020xvf} the corresponding estimates of Ref. \cite{SinghSidhu:2020cxw} are oversimplified and do not have the same status as those derived from CMB or BBN constraints.  
}  
Assuming the conventional halo model with a local DM density of $\rho_{\rm DM}\simeq 0.3\,{\rm {GeV} {cm^{-3}}}$, the rate at which AQNs hit the Earth \cite{Lawson:2019cvy} is
\be
\label{eq:D Nflux 3}
\frac{\langle\dot{N}\rangle}{4\pi R_\oplus^2}
&\simeq & \frac{4\cdot 10^{-2}}{\rm km^{2}~yr}
\left(\frac{\rho_{\rm DM}}{0.3{\rm \frac{GeV}{cm^3}}}\right)
 \left(\frac{\langle v_{\rm AQN} \rangle }{220~{\rm \frac{km}{s}}}\right) \left(\frac{10^{25}}{\langle B\rangle}\right),\nonumber \\
 \langle\dot{N}\rangle&\simeq & 0.67\,{\rm s}^{-1} \left(\frac{10^{25}}{\langle B\rangle}\right)\simeq  2.1\cdot 10^7 {\rm yr}^{-1} 
\left(\frac{10^{25}}{\langle B\rangle}\right).
\ee
In obtaining  Eq.~\ref{eq:D Nflux 3} it was assumed that $\langle\dot{N}\rangle \approx 4\pi b^2 n_{\rm AQN}v_{\rm AQN}$, where $b\approx 1.0013 R_{\oplus}$ is the impact parameter for Earth, and we averaged over incident angles of impacting AQNs; see Appendix C in Ref.~\cite{Lawson:2019cvy} with details. We also assume that the  AQNs represent the dominant portion of the DM. Numerical expressions for Eq.~\ref{eq:D Nflux 3} were derived in Ref.~\cite{Lawson:2019cvy} using full-scale Monte Carlo simulations which account for all types of AQN trajectories with different AQN masses $ M_N\simeq m_p|B|$, different incident angles, and different initial velocities and size distributions. It was shown that none of these factors significantly affects our estimate. The result in Eq.~\ref{eq:D Nflux 3} suggests that AQNs with a typical size $B \approx 10^{25}$ hit the Earth's surface with a frequency on the order of once a day (i.e., hundreds per year)  per $100\times100\,{\rm km^2}$.

We conclude this brief review subsection with Table \ref{tab:basics}, which summarizes the fundamental features and parameters of the AQNs. The parameter $\kappa$ in Table \ref{tab:basics} accounts for the fact that not all matter interacting with the nugget will annihilate, and not all of the energy released during the annihilation will be thermalized within the nuggets. The ratio $\Delta B/B \ll 1$ in the Table ~\ref{tab:basics} indicates that only a small portion, $\Delta B$, of the total (anti)baryon charge $B$ hidden in the form of AQNs gets annihilated during BBN, CMB, or post-recombination epochs (including galaxy and star formation), while the majority of the baryon charge survives to the present day. Independent analyses \cite{Santillan:2020lbj} and \cite{SinghSidhu:2020cxw} further support our original claims, as cited in Table \ref{tab:basics}, that the antiquark nuggets survive the BBN and CMB epochs.

 \exclude{
Finally, the AQN model with the same set of parameters may explain several other puzzling observations in different environments. In particular, it may explain the primordial Li puzzle \cite{Flambaum:2018ohm} during the early Universe evolution. On a different, galactic scale, it may explain the observed excess of the UV radiation \cite{Zhitnitsky:2021wjb}. The excess of radiation is observed in many different frequency bands, from UV to optical, IR, and even radio frequency bands. It is normally expressed as excess in the Extra-galactic Background Light (EBL), see e.g. review \cite{Mattila:2019ybk}. The energy injection due to the AQN-induced processes may explain a portion or even entire observed excesses in different frequency bands.

It may also explain some observed inconsistencies on small galactic scales (such as the Core-Cusp problem \cite{Flores-1994-observational, Moore-1994-evidence}, the Too-Big-to-Fail problem \cite{Boylan-2011-too-big}, etc) which are hard to explain within the conventional CDM paradigm \cite{Zhitnitsky:2023znn}. It may also shed some light on the long-standing solar corona heating problem \cite{Zhitnitsky:2017rop, Raza:2018gpb}, which could be the manifestation of the AQN's processes at different scales (outside of the galactic scale where the DM is expected to manifest itself). Finally, the AQNs may also explain some mysterious events observed on the Earth scale, such as the anomalous events recorded by ANITA \cite{Liang:2021rnv}.    
}
 
\section{AQN-induced axions}
\label{AQN-induced}

As previously noted, AQNs traveling through the Earth's atmosphere and interior lose baryon charge due to annihilation processes, which in turn lead to axion emission as proposed in Ref.~\cite{Fischer:2018niu}. This novel axion production mechanism can be outlined as follows: the total energy of an AQN reaches an equilibrium minimum when the axion domain wall contributes roughly one-third of the AQN’s total mass \cite{Ge:2017idw}. In this equilibrium state, the configuration does not emit axions due to a purely kinematic constraint; the static axion domain wall can be considered a superposition of off-shell, nonpropagating axions. However, this static balance is disrupted when a baryon charge is annihilated through interactions with the environment, causing time-dependent perturbations that shift the AQN out of equilibrium. In essence, the AQN configuration becomes unstable and prone to axion emission because it no longer resides at its minimum energy with fewer baryon charges in the quark nugget core. As annihilation proceeds, the AQN gradually loses mass, leading the axion domain wall around the quark matter core to oscillate and contract. These domain wall oscillations generate excitation modes that eventually result in the emission of propagating axions. The spectrum for these AQN-induced axions has been calculated in Ref.~\cite{Liang:2018ecs}.
    
Several key features of this spectrum merit mention. First, the typical velocities $\la v_a^{\rm AQN}\ra \simeq 0.6 c$ are significantly higher than those of conventional galactic nonrelativistic axions, which have $\la v_a\ra \sim 10^{-3} c$. Monte Carlo simulations presented in Ref.~\cite{Liang:2019lya} provide the following estimate for the density of these relativistic axions at the Earth’s surface:
\be
\label{denisty_relativitic}
\la \rho_\mathrm{a}^{\rm AQN}(R_\oplus)\ra \sim 5\cdot 10^{-6}A(t){\rm  \frac{GeV}{cm^3}}, ~ \la v_\mathrm{a}^{\rm AQN}(R_\oplus)\ra \simeq 0.6 c,~~~~
\ee
where $A(t)$ is the time-dependent modulation/amplification factor. The factor $A$ for the daily and annual modulations is discussed in Sec.~\ref{strategy} below, specifically in Eq.~\ref{eq:daily} and \ref{eq:annual}. In both cases, $A$ does not deviate from the average value $\la A\ra=1$ by more than $10\%$. 

However, this normalization assumes the Standard Halo Model (SHM) distribution for DM, with a density of approximately $\rho_{\rm DM} \approx 0.3 ~{\rm GeV}/{\rm cm^3}$. The actual DM density on the Earth’s surface could differ significantly from this standard value due to the gravitational capture of DM particles by stars and planets; see, for example, Ref.~\cite{Leane:2022hkk} and references therein. Indeed, one of the primary motivations in Ref.~\cite{Liang:2020mnz} was to critically examine the SHM assumption within the AQN framework through specific computations.

The AQN-induced axions are predominantly generated in the Earth’s deep interior, where the density of surrounding matter is highest. Notably, the energy density of the axions [Eq.~\ref{denisty_relativitic}] and the energy flux [Eq.~\ref{flux}] are relatively insensitive to the axion mass $m_a$, as detailed in Ref.~\cite{Liang:2019lya}. This behavior contrasts with that of galactic nonrelativistic axions, under the assumption that DM is primarily composed of these galactic axions.
\exclude{
\footnote{\label{saturation}The assumption on saturation of the DM density by the galactic axions cannot be satisfied in the entire window of $ 10^{-6} {\rm eV} \lesssim m_a\lesssim 10^{-3} {\rm eV}$ as the conventional contribution is highly sensitive to $m_a$ as $\rho_{a} \sim m_a^{-7/6}$ and may saturate the DM density at $m_a\lesssim 10^{-5} {\rm eV}$, depending on additional assumptions on production mechanism.  This, in particular, implies that for $m_a\gtrsim 10^{-4} {\rm eV}$ the conventional galactic axions contribute very little to $\Omega_{\rm DM}$, while the AQNs may be the dominant contributor to the DM density irrespective to the axion mass $m_a$.}.
}

It is useful to compare the energy density of the relativistic AQN-induced axions [Eq.\ref{denisty_relativitic}] with that of CMB photons, as both are relativistic components. The energy density of CMB photons is well established at $\rho_{\gamma} \approx 0.26{\rm eV/cm^3}$, nearly 4 orders of magnitude lower than the AQN-induced axion density on Earth’s surface (Eq.~\ref{denisty_relativitic}). By contrast, the density of CaB axions discussed in Ref.~\cite{Dror:2021nyr} and referenced in Sec.~\ref{sect:introduction} must satisfy $\rho_{\rm CaB} \ll \rho_{\gamma}$. This constraint implies that the CaB axion density is at least four orders of magnitude lower than the density of AQN-induced axions at Earth’s surface [Eq.~\ref{denisty_relativitic}].

There is a significant difference in the distribution of CMB photons compared to the AQN-induced axion energy density. CMB photons are nearly uniformly distributed across the Universe, while the axion density [Eq.~\ref{denisty_relativitic}] is concentrated around the Earth and rapidly decreases at distances much greater than $R_\oplus$. Similar localized distributions can be estimated for other stars and planets. Unfortunately, axions interact far more weakly with visible matter than CMB photons, which poses a substantial challenge for their detection despite their higher density near the Earth (Eq.~\ref{denisty_relativitic}) relative to CMB photons.

The resulting number density $n_a^{\rm AQN}\simeq \rho_\mathrm{a}^{\rm AQN}/m_a$  is approximately 5 orders of magnitude smaller than the conventional galactic axion number density assuming that the galactic nonrelativistic axions saturate the DM today. However, the flux $(v_\mathrm{a}^{\rm AQN}\cdot n_a^{\rm AQN})$ related to relativistic axions (Eq.~\ref{denisty_relativitic}) is only 2 orders of magnitude smaller than the conventional flux of nonrelativistic axion. Indeed, the AQN-induced axion energy flux on the Earth's surface can be estimated as follows \cite{Liang:2019lya}:
\be
\label{flux}
\la E_a \cdot \Phi^{\rm AQN}_a  \ra \simeq 10^{14} A(t) \left[{\rm\frac{eV}{cm^2\cdot s}}\right], ~~~ \la E_a\ra\simeq 1.3\,m_a.
\ee
It is instructive to compare the AQN-related flux [Eq.~\ref{flux}] with the flux computed from the assumption that the galactic axions saturate the DM density $\rho_{DM}\sim 0.3\,{\rm GeV\cdot cm^{-3}}$ today. In this case, the numerical value for the flux [Eq.~\ref{flux}] is approximately 2 orders of magnitude below the value $10^{16} {\rm {eV}/({cm^2\cdot s})} $ computed for the conventional galactic nonrelativistic axions assuming these axions saturate the DM density. It is emphasized once more that the density [Eq.~\ref{denisty_relativitic}] as well as the flux [Eq.~\ref{flux}] do not strongly depend on the axion mass $m_a$ nor the initial misalignment angle $\theta_0$ in contrast with the conventional expression for the galactic axions. 

 \exclude{
It is noted that the factor $A(t)$ could be numerically very large for a very short period in case of rare bursts-like events, the so-called ``local flashes'' in the terminology of Ref.\,\cite{Liang:2019lya}. These short bursts  (with a duration time of the order of a second for $A\simeq 10^2$) result from the interaction of the AQN hitting the Earth in the close vicinity of a detector, see Table \ref{tab:local flashes}.

\begin{table} 
\captionsetup{justification=raggedright}
\caption{Estimations of Local flashes for different $A$ as defined by Eq.~\ref{flux}. The corresponding event rate and the time duration $\tau$ depend on the factor $A$, which is determined by the shortest distance from the nugget's trajectory to the detector. The table is adopted from \cite{Liang:2019lya}: } 
	\centering 
	\begin{tabular}{ccc}
		\hline \hline
		$A$ &  $\tau$ (time span) & event rate \\ 
		\hline 
		1 & 10 s & 0.3 $\rm min^{-1}$ \\ 
		$10$ & 3 s & 0.5 $\rm hr^{-1}$ \\ 
		$10^2$ & 1 s & 0.4 $\rm day^{-1}$ \\ 
		$10^3$ & 0.3 s & 5 $\rm yr^{-1}$ \\  
		$10^4$ & 0.1 s & 0.2 $\rm yr^{-1}$ \\  
		\hline 
	\end{tabular}
	\label{tab:local flashes}
\end{table}
}

A key characteristic of AQN-induced axions is their daily modulation, a property not typically found in generic DM models, especially when detectors lack sensitivity to directionality, as is common with most modern DM detectors. This daily modulation is a distinctive feature of the AQN-induced production mechanism and is central to our analysis.

The origin of this modulation lies in the difference in AQN size between its entry and exit points, resulting from annihilation processes within the Earth's interior.
\exclude{\footnote{On average a typical nugget loses approximately $30\%$ of its baryon charge when crosses the Earth. The axion domain wall shrinks correspondingly, which eventually generates the axion density (Eq.~\ref{denisty_relativitic}) at the Earth's surface from all AQNs trans-passing the Earth's interior.}.}
Such effects are absent for fundamental particles like wealy interacting massive particles (WIMPs), whose mass remains unchanged between entry and exit. In contrast, the average size difference for AQNs as they enter and exit Earth’s surface can produce a significant daily modulation effect, approximately 10\%, as calculated in Ref.~\cite{Liang:2019lya} and discussed below.

\begin{figure}
    \centering
    \includegraphics[width=0.8\linewidth]{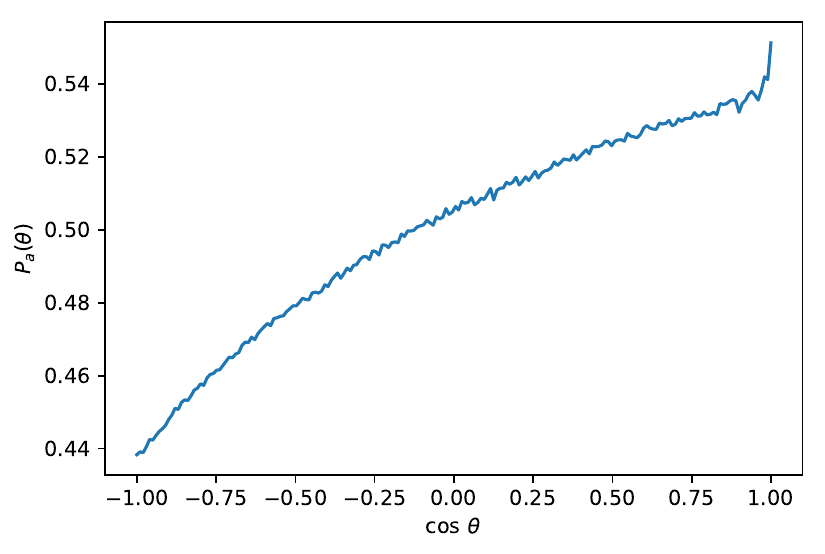}
    \caption{The result of the Monte Carlo numerical simulations for the azimuthal distribution of axion flux $P_a(\theta)$ on the surface of the Earth, adopted from \cite{Liang:2019lya}. A relative orientation of the Earth's axis of rotation and DM galactic wind generates the flux asymmetry sensitive to the latitude $\theta$, which consequently leads to the daily modulation effect.}
    \label{fig:Pa_theta}
\end{figure}

The key factor driving the daily modulation here is the relative orientation between Earth's rotational axis and the galactic DM wind. Unlike WIMPs, where this orientation is irrelevant without directional detectors, AQN fluxes are sensitive to it, generating a flux asymmetry that varies with latitude, $\theta$, and leads to daily modulation. This feature of the axion flux is captured by the parameter $P_a(\theta)$, which is normalized such that $\int P_a(\theta)\sin\theta\, d\theta = 1$. In the absence of daily modulation, $P_a(\theta)$ would be uniform at 0.5, independent of latitude. However, our numerical Monte Carlo simulations, shown in Fig.~\ref{fig:Pa_theta}, clearly demonstrate a $\theta$ dependence that produces daily modulation.

It is worth noting that the daily modulation discussed here differs substantially from that in Ref.~\cite{ADMX:2023rsk}, where the directional sensitivity of the instrument plays a central role in detecting daily modulations from CaB. In our case, even a spherically symmetric detector, which lacks directional sensitivity, can still measure this daily modulation effect.

The production rate of the low-energy AQN-induced axions with $v_a\ll c$ is strongly suppressed as explained above. However, the axions that are produced with extremely low velocities $v_a\lesssim 11 \rm ~ {km}/{s}$ will be trapped by the Earth's gravitational field. These axions will be orbiting the Earth indefinitely; therefore, they will be accumulated around the Earth during its lifetime of 4.5 billion years.
 
The corresponding Monte Carlo simulations have been performed in Ref.~\cite{Lawson:2019cvy} with the following estimate for the gravitationally bound axions:
\be
\label{denisty_bound}
\rho_\mathrm{a}^{\rm bound}(R_\oplus)\sim 10^{-4}{\rm  \frac{GeV}{cm^3}}, ~~~ \la v_\mathrm{a}(R_\oplus)\ra \simeq 8 \rm \frac{km}{s}.
\ee
The number density of the bound axions is at least 2 orders of magnitude smaller than conventional axion number density assuming that the galactic nonrelativistic axions saturate the DM density. However, the corresponding wavelength $\lambda_\mathrm{a}\sim {\hbar}/({m_\mathrm{a} v_\mathrm{a}})$ of the gravitationally bound axions is approximately $30$ times greater than for galactic axions, which have a typical velocity of about $\sim 250  $ km/s. Therefore, coherent effects can be maintained for a longer time period compared to those for conventional galactic axion searches. One may hope that the feature of having a large coherence length, $\lambda_\mathrm{a}\sim v_\mathrm{a}^{-1}$, could play a key role in the design of instruments, capable of discovering such gravitationally trapped axions. 

These AQN-induced axions (relativistic as well as gravitationally bound) can be treated as a classical field because the number of the AQN-induced axions (Eq.~\ref{denisty_relativitic}) accommodated by a single de Broglie volume is very large even though the de Broglie wavelength $\lambda$ for relativistic AQN-induced axions is much shorter than for galactic axions, 
 \be
\label{density}
n_a^{\rm AQN}\lambda^3\sim \frac{\la \rho_\mathrm{a}^{\rm AQN}\ra}{m_a} \left(\frac{\hbar}{m_a v_a}\right)^3\sim 10^6 \left(\frac{10^{-4} {\rm eV}}{m_a}\right)^4\gg 1, \nonumber
 \ee
where we use $m_a\sim 10^{-4} \rm eV$ as a benchmark axion mass to be used for all numerical estimates in the present work.

\section{New strategy: detection of broadband axions}
\label{strategy}

The large average velocities $\la v_a\ra \simeq 0.6 c$ of the emitted axions by AQNs change the entire strategy of axion searches since they are characterized by a broad distribution with $m_a\lesssim \omega_a\lesssim 1.8~m_a$ as discussed in the previous section. Therefore, the corresponding axion detectors must be some kind of broadband instrument. The cavity-type experiments such as ADMX \cite{ADMX_2018_search} and CAPP \cite{ahn_20224_extensive} are to date the only ones to probe the parameter space of the conventional QCD axions with $\la v_a\ra \sim 10^{-3} c$, while we are interested in the detection of the relativistic axions with $\la v_a\ra \sim 0.6 c$. This requires different types of instruments and drastically different search strategies. We assume that some kind of broadband instruments can be designed and built, see reviews \cite{Marsh:2015xka, Graham:2015ouw, Battesti:2018bgc, Irastorza:2018dyq, Sikivie:2020zpn} for a description of possible detectors. 

With this assumption in mind, a new strategy to probe the QCD axion can be formulated as follows \cite{Budker:2019zka}. It has been known since Ref.~\cite{Freese:1987wu} that the DM flux shows annual modulation due to the differences in relative orientations of the DM wind and the direction of the Earth's motion around the Sun. The corresponding effect for AQN-induced axions was computed in Ref.~\cite{Liang:2019lya}. The daily modulation which is a very specific feature of the AQN model, as explained in the previous section, was also computed in the same paper \cite{Liang:2019lya}. The broadband strategy is to separate a large frequency band into a number of smaller frequency bins with the width $\Delta \nu \sim \nu$ according to the axion dispersion relation as discussed above. 
 
The time-dependent signal in each frequency bin $\Delta \nu_i$ has to be fitted according to the expected modulation pattern, daily or annual. For example, the daily modulation should be fitted according to the formula 
\be 
\label{eq:daily}
&&\rho_\mathrm{a}^{\rm AQN}(t) \equiv A_{\rm (d)}(t) \la \rho_\mathrm{a}^{\rm AQN}(R_\oplus)\ra \\  \nonumber
&&A_{\rm (d)}(t)\equiv[1+\kappa_{\rm (d)} \cos(\Omega_d t-\phi_0)],
\ee
where the density $ \la \rho_\mathrm{a}^{\rm AQN}(R_\oplus)\ra$ is estimated according to Eq.~\ref{denisty_relativitic}.
In this formula $\Omega_d=2\pi\,\rm day^{-1} $ is the angular frequency of the daily modulation, while $\phi_0$ is the phase shift. It can be assumed to be constant on the scale of days. However, it slowly changes during the annual seasons due to the variation of the direction of DM wind with respect to the Earth. The modulation factor $A_{\rm (d)}(t)$ is normalized to unity, $\la A_{\rm (d)}(t)\ra =1$ when averaged over a large number of daily cycles. The same modulation factor $A_{\rm (d)}(t)$ with the same $\kappa_{\rm (d)}$ also enters the estimation for the flux [Eq.~\ref{flu]}]. 

A similar formula holds for the annual modulation defined as:   
\be 
\label{eq:annual}
 && \rho_\mathrm{a}^{\rm AQN}(t) \equiv A_{\rm (a)}(t) \la \rho_\mathrm{a}^{\rm AQN}(R_\oplus)\ra \\ \nonumber
 &&A_{\rm (a)}(t)\equiv[1+\kappa_{\rm (a)} \cos\Omega_a (t-t_0)], 
\ee
where $\Omega_a=2\pi\,\rm yr^{-1} $ is the angular frequency of the annual modulation and label $``a"$ in $\Omega_a $ stands for annual.   The $\Omega_a t_0$ is the phase shift corresponding to the maximum on June 1 and minimum on December 1 for the standard galactic DM distribution, see Refs.~\cite{Freese:1987wu, Freese:2012xd}. 
 
Several tests can unambiguously discriminate a true genuine signal from spurious signals and noise. These tests will be discussed in detail in Sec.~\ref{sec:modulation} after some simplified estimates for the expected signal are performed in Sec.~\ref{sec:estimates}.

\section{Expected Signals from the AQN-induced axions}
\label{signals}

The goal of this section is twofold. First, in Sec.~\ref{sec:estimates} we want to highlight the basic differences in the conversion rate between the AQN-induced relativistic axions and the conventional galactic axions which normally enter the well-known formulae. Second, in Sec.~\ref{sec:modulation} we formulate the key element of this work: despite the very low rate of axion-photon conversion the daily modulation effect reviewed in the previous Sec.~\ref{strategy} effectively allows us to discriminate the genuine signal from the spurious signals. 

\subsection{Estimates for the axion to photon conversion}
\label{sec:estimates}

The starting point for our analysis is the computation of the conversion rate which is determined by the well-known formula (e.g. see Ref.~\cite{Sikivie:2020zpn} for a review):
\be
\label{conversion}
&&P_{a\rightarrow \gamma}= \frac{1}{v_a}\left(\frac{g\cal{B}}{q}\right)^2\sin^2\left(\frac{qL}{2}\right), \\
 &&q=\omega-\sqrt{\omega^2-m_a^2} \nonumber
 \ee
where $L$ is a typical distance where the magnetic field $\cal{B}$ is  present and $g\equiv g_{a\gamma\gamma}$ is defined as usual
\be
g\simeq  \frac{g_{\gamma}\alpha}{f_a\pi}, ~~ g_{\gamma}=\frac{1}{2}\left (\frac{N_e}{N}-\frac{5}{3}- \frac{m_d-m_u}{m_u+m_d} \right) 
 \ee
where the parameter $N_e/N=0$ for KSVZ model, and  $N_e/N=8/3$ for DFSZ model such that $g_{\gamma}\approx -0.97$ for the Kim-Shifman-Vainshtein-Zakharov (KSVZ) \cite{KSVZ1,KSVZ2} model and 
$g_{\gamma}\approx 0.36$ for the Dine-Fischler-Srednicki-Zhitnitsky (DFSZ) \cite{DFSZ1,DFSZ2} model. The parameters $f_a$ and $m_a$ are not independent, but tightly linked in conventional QCD axion models:
 \be
 \label{f_a}
 m_a\approx 6\cdot 10^{-4}  \left(\frac{10^{10} \rm ~GeV}{f_a} \right) {\rm eV}.
 \ee
  
In case of the resonant conversion when combination  $(qL)\ll 1$  could be very small the conversion rate assumes the usual form $P_{a\rightarrow \gamma}=v_a^{-1}(g {\cal{B}} L/2)^2$. The dependence $P_{a\rightarrow \gamma}\propto L^2$ should be interpreted as the presence of a coherent phenomenon when the amplitude of the transition is linearly proportional to the coherence length $\propto L$, while the probability  $P_{a\rightarrow \gamma}$ shows the quadratic dependence $P_{a\rightarrow \gamma}\propto L^2$ as stated above. 

For nonrelativistic axions, which are normally assumed to be the case for haloscope studies, we can approximate $q \simeq m_a$. If Eq.~\ref{conversion} is multiplied with the axion flux $(\rho_av_a)/m_a$ we arrive at the following expression for the flux $\Phi_{\gamma}$ which counts the number of microwave photons emerging from the magnetic field region per unit time and unit area:
\be
\label{gamma-flux}
\Phi_{\gamma}=\frac{(\rho_a v_a)}{m_a}P_{a\rightarrow \gamma}\approx \frac{\rho_a}{m_a}\left[ \frac{g {\cal{B}}}{m_a}\sin\left(\frac{m_aL}{2}\right)\right]^2.
\ee 
This expression agrees with the formula of ref. \cite{Ioannisian:2017srr} which was derived using a very different approach. It is important to note that Eq.~\ref{gamma-flux} does not depend on the axion's velocity $v_a$ in the nonrelativistic limit $v_a\rightarrow 0$. This is because Eq.~\ref{conversion} for the conversion is proportional to $v_a^{-1}$ while computation of the $\Phi_{\gamma}$ requires multiplication of the conversion $P_{a\rightarrow \gamma}$ to the axion flux which itself is proportional to $v_a$ such that the dependence on $v_a$ cancels out at small $v_a/c\ll 1$.

We now turn to our relativistic case for the AQN-induced axions. In this case, the axion density and the flux are determined by Eq.~\ref{denisty_relativitic} and Eq.~\ref{flux} correspondingly. Furthermore, the time formation of a photon is very short $\sim m_a^{-1}$. The corresponding length scale is determined by the de Broglie length $\lambda\sim m_a^{-1}$ which is also much shorter than the size of the system $L$. Therefore, the conversion process can be treated as it happens in an infinite volume system.  

As a result, the flux $\Phi^{\rm AQN}_{\gamma}$  can be estimated, which counts the number of microwave photons emerging from  the magnetic field region per unit time and unit area as a result of a {\it single} elementary process as:
\be
\label{gamma-AQN}
\Phi^{\rm AQN}_{\gamma}\approx \Phi^{\rm AQN}_a P_{a\rightarrow \gamma}\approx  \Phi^{\rm AQN}_a\left[ \frac{g {\cal{B}}}{m_a}\sin\left(\frac{m_aL}{2}\right)\right]^2,~~
\ee 
Here, we neglect numerical factors of order of 1 and assume $\la E_a\ra \simeq m_a, ~\la v_\mathrm{a}^{\rm AQN}\ra \simeq 1$ for simplicity. We also disregard complications related to the relative orientation of the magnetic field and axion momentum, as well as differences between $L_x, L_y, L_z$ that characterize the magnetic field geometry. All these effects are negligible for our qualitative analysis, given that the detector $L$ in all directions is much larger than the Compton wavelength $m_a^{-1}$, which determines the typical length scale of an elementary process, ensuring $(m_aL)\gg 1$. This should be contrasted with the canonical treatment of the problem when the de Broglie wavelength is very large, i.e. $\lambda\sim (m_a v_a)^{-1}\gg L$, in which case the geometrical properties of the cavity become important and a single cavity mode gets excited. In contrast, in the relativistic case, many modes get excited simultaneously with approximately the same quality factor of order 1. Therefore, higher-order modes, if excited by AQNs, can contribute to the total received power. However, in practice, the coherent addition of the power of the higher-order modes is non-trivial, as the individual phases of these modes must be carefully accounted for. This complexity underscores the challenges in detecting broadband signals in resonant cavities. Further comments on this issue are provided in item 5 of Sec.~\ref{conclusion}.

In principle, we should integrate over the axion spectrum computed previously with a specific geometry of the instrument when  $E_a$ approximately varies in the range  $E_a\in (m_a-1.8 m_a)$. The non-resonant Primakoff conversion generates a complicated photon's broadband spectrum with $\nu\in (m_a-1.8 m_a)$. However, for our qualitative estimates when we want to study the parametrical dependence of the rate as a function of the key parameters such as the axion mass $m_a$, magnetic field ${\cal{B}}$, the basic size of the detector $L$, all these numerical factors of the order of one are irrelevant and will be ignored in what follows. 
        
Several key elements lead to the different treatments of the nonrelativistic galactic axions in comparison with relativistic AQN-induced axions. First of all, the conventional treatment of the problem assumes the coherent conversion such that $P_{a\rightarrow \gamma}\propto L^2$ as explained above. This $L^2$ scaling holds as long as $L$ is smaller than the De Broglie wavelength $\lambda$ where conversion effectively occurs. If coherence persists up to distance $l<L$ the enhancement is much smaller, $L^2\rightarrow Ll$ \cite{Sikivie:2020zpn}. It should be contrasted with the AQN induced axions when $\Delta \nu/\nu\sim 1 $ and one should average over many cycles such that $\sin^2(\frac{qL}{2})\rightarrow 1/2$ in Eq.~\ref{gamma-AQN}. 
        
The second important element is as follows. Equation \ref{gamma-AQN} describes the conversion as a result of a {\it single} elementary process when a typical formation length is of order the Compton length $m_a^{-1}$. One should sum over the number of effective layers along the axion's path which is of order $(Lm_a)\gg 1$. This factor represents a strong enhancement factor. It is noted that this factor  $(Lm_a)\gg 1$ agrees with formula $Ll$  presented in Ref.~\cite{Sikivie:2020zpn} when the coherence length $l$ should be replaced to the Compton length, i.e., $l\rightarrow  m_a^{-1}$. 
        
Therefore, based on the above, we arrive at the following estimate for the total flux $\Phi^{\rm AQN}_{\gamma}$ which counts the number of microwave photons emerging from   the magnetic field region per unit time and unit area:
\be
\label{AQN-rate}
\Phi^{\rm AQN}_{\gamma} ({\rm tot.})\approx \frac{A}{2}\Phi^{\rm AQN}_a\cdot \left[ \frac{g {\cal{B}}}{m_a}\right]^2\cdot (m_aL). 
\ee      
Numerically, this estimate can be presented as: \footnote{We use the conversion factor for the magnetic field when $\rm Gauss\approx 1.95 \cdot 10^{-2} {\rm eV^2}$ in Heaviside-Lorentz units when $\alpha^2=e^2/(4\pi)$, see e.g. Appendix A in \cite {Sikivie:2020zpn}.}:
\be
  \label{numerics}
  \Phi^{\rm AQN}_{\gamma} ({\rm tot.})\approx 0.1A \left(\frac{g_{\gamma}}{0.97}\right)^2\left(\frac{{\cal{B}}}{10 T}\right)^2\left(\frac{L}{m}\right)    {\rm\frac{ ph.}{m^2   day}},  ~~~~
\ee
which implies that a detector with area  $\sim 10~m^2$ could see a few microwave photons a day. This rate, of course, is quite low. However, recent technological advances suggest that such photons can be, in principle, detected using   such instruments as 
Broadband Reflector Experiment for Axion Detection (BREAD). We refer to Ref.~\cite{BREAD:2021tpx} on broadband technology in axion searches.

    \exclude{
What is important here is that this rate could get a significant boost for a short period of time when factor $A(t)$ could be much larger (than an average value $\la A\ra=1$) resulting from the interaction of the AQN hitting the Earth in close vicinity of a detector, see Table \ref{tab:local flashes} with the expected frequency of possible enhancements. For example, one should expect a daily burst with enhancement $A\sim 10^2$ which lasts for a second as shown in Table \ref{tab:local flashes}. The rarer bursts are shorter in duration but could be much more powerful and pronounced.
    }
    
Another point here is that the average rate is low, however, no detector tuning is required in contrast with conventional resonant-type experiments (see, for example, Ref.~\cite{Semertzidis_2022_axion} for a review). Furthermore, as we already mentioned, the AQN-induced axion DM flux [Eq.~\ref{flux}] does not depend on the axion mass $m_a$ in contrast with the conventional axion production mechanisms when $\rho_{\rm DM} \sim m_a^{-7/6}$. As a result, the rate in Eq.~\ref{numerics} also does not depend on $m_a$ and this holds as long as our assumption on relatively large mass is satisfied, i.e. $(m_aL)\gg 1$.

The conversion rate in Eq.~\ref{conversion} allows us to estimate the excess of power due to the emitted microwave photons. First, we have to multiply the axion density [Eq.~\ref{denisty_relativitic}] by the conversion rate in Eq.~\ref{conversion}. Second, an average over many cycles should be performed such that $\sin^2(\frac{qL}{2})\rightarrow 1/2$ and insert the enhancement factor $(Lm_a)\gg 1$ as explained after Eq.~\ref{gamma-AQN}. Third, to estimate the released energy per unit of time, the obtained expression should be divided to the typical formation time of the elementary process, $\Delta t \sim m_a^{-1}$. Collecting all these factors together we arrive at the following estimate for the power excess $P^{\rm AQN}_{\gamma} ({\rm tot.})$ due to the emitted microwave photons in unit volume:
   \be
\label{AQN-power}
P^{\rm AQN}_{\gamma} ({\rm tot.})\approx    A(t)\left(\frac{\la\rho^{\rm AQN}_a\ra m_a }{2}\right)  \left( \frac{g {\cal{B}}}{m_a}\right)^2 (m_aL). ~~~~
\ee    
This estimate can also be represented in a conventional way as an excess of power in unit volume measured in $W/m^3$:
\be
  \label{numerics1}
 P^{\rm AQN}_{\gamma} ({\rm tot})&\approx&  10^{-26} A(t) \left( {\rm \frac{Watt}{m^3}}  \right)\left(\frac{g_{\gamma}}{0.97}\right)^2 \\
 &\times& \left(\frac{{\cal{B}}}{10 ~\rm Tesla}\right)^2 \left(\frac{L}{m}\right)  \left(\frac{m_a}{10^{-4} \rm eV}\right)^2.    \nonumber
\ee
The theoretical estimate (\ref{numerics1}) does not include additional factors that may considerably modify the numerical value of $10^{-26}$. First of all, as we already mentioned we do not distinguish the geometrical parameters  $L_x, L_y, L_z$ which we assume are similar in values. Another factor that may potentially modify a numerical value of (\ref{numerics1}) is the different sensitivity of a cavity to specific frequency bands, the ``filters". We do not include such factors in the estimate (\ref{numerics1}) as these modifications could be very different for different instruments, while (\ref{numerics1}) represents a very generic parametrical behavior for an excess of power as a function of $m_a$ describing the conversion of the relativistic axions. In particular, it includes the most important features of the system such as the typical formation time of the elementary process $\sim m_a^{-1}$ and the presence of a large number of layers along the path of relativistic axions. It is noted, that the value of $m_a \sim 10^{-4} \, \text{eV}$ serves as a benchmark value for all our numerical estimates in the present work as mentioned at the end of Sec. \ref{AQN-induced}.  As long as the condition $(m_a L) \gg 1$ is satisfied, our assumptions remain valid, and our parametric dependence on $m_a$ holds.

It is instructive to compare this estimate with the conventional computations for the power excess $P_a$ in cavity resonant-designed detectors when $P_a\propto 10^{-26} Q_{\alpha} \rm (W/m^3)$, see \cite{Sikivie:2020zpn} for a review. Numerically, the value presented in \cite{Sikivie:2020zpn} is similar to our estimate (Eq.~\ref{numerics1}) though the conversion rates and the magnitudes for the axion densities of the AQN-induced and the galactic axions are very different. This numerical similarity will be used in what follows for illustrative purposes in our comparisons of the broadband strategy with conventional resonant type searches. 

The key element in resonant searches is the so-called quality factor $Q_{\alpha}\sim 10^6$, which increases the likelihood of the detection of the axion signal. Indeed, the power excess $P_a$ in the conventional analysis is normally compared with the noise power $P_{\rm noise} $ such that the signal-to-noise ratio is determined by the formula
\be
\label{SN-ratio}
\frac{S}{N}\approx\frac{ \sqrt{N}P_a}{P_{\rm noise}}\approx \frac{P_a}{T}\sqrt{\frac{t}{\Delta \nu}},~ P_{\rm noise} =
T\Delta \nu, ~ N\approx t\Delta \nu \nonumber
\ee
where $T$ is the total system noise temperature, $t$ is the integration time  and  the $\Delta \nu$  is the bandwidth of the axion signal. The ``quality factor" $Q_{\alpha}\sim 10^6$ plays a crucial role in this approach. The basic idea here is to detect the narrow resonance line with  $\Delta \nu\simeq 10^{-6} \nu$ with sufficiently long integration time by scanning each narrow frequency band to increase the signal-to-noise ratio. 

However, the formula for the power excess $P^{\rm AQN}_{\gamma} ({\rm tot})$, as given by Eq.~\ref{numerics1}, does not carry the enhancement factor $Q_{\alpha}\sim 10^6$  for the case of non-resonant conversion of the relativistic axions with $\Delta \nu\sim \nu$. Therefore, the entire logic of collecting the signal and discriminating it from the noise is very different from the conventional treatment of the cavity-type experiments briefly reviewed above. The key point for our studies is that there is a well-defined modulation factor $A(t)$ which allows us to discriminate the signal from the spurious events and noise by collecting the signal over entire frequency bands instead of scanning in a single narrow frequency band in cavity-type experiments. In particular, a daily modulation is expected, as discussed in the previous section.

The absence of the quality factor $Q_{\alpha}\sim 10^6$ in Eq.~\ref{numerics1} implies that the broadband search is not particularly sensitive to the axion mass, except for a basic normalization factor $\sim m_a^2$ resulted from a large number of ``layers" along the path of the relativistic axion, as explained above. This should be contrasted with conventional resonant type searches when the cavity is designed to be highly sensitive to a very specific narrow frequency band expressed in terms of $Q_{\alpha}$. This feature (absence of the resonant sensitivity to $m_a$)  is of course a direct manifestation of the non-resonant axion-photon conversion in the magnetic field when the resonant features of the cavity do not manifest themselves in the expression of the power excess as given by Eq.~\ref{numerics1}. 

Therefore, the only option we are aware of to determine the axion mass $m_a$ in the broadband setting is to determine the frequency of a single photon recorded by a non-resonant axion photon converter. As mentioned after Eq.~\ref{numerics} such technologies that could detect a single photon in this frequency band may indeed become available soon.   

We would like to illustrate the arguments presented above with the following numerical estimates. In conventional resonance-type searches the factor $N\approx t\Delta \nu\approx 10^5$ for  $\Delta \nu\simeq 10^{-6} \nu$ and $\nu\sim \rm 1~GHz$ such that $\delta P_{\rm noise} \approx P_{\rm noise}/\sqrt{N}\approx 0.3\cdot 10^{-2}P_{\rm noise}$ for $t\approx 1$ min which is a typical integration time in resonant searches. 

It should be contrasted with broadband strategy when a signal is collected over the entire frequency band $\Delta \nu\sim \nu\sim \rm 1~GHz$ during the entire year (separately for each hour during the day). In this case the integration time $t\approx 10^6$ s and $N\approx t\Delta \nu\approx 10^{15}$ which is more than sufficient to overcome the absence of the quality factor $Q$ in the broadband Eq.~\ref{AQN-rate}, Eq.~\ref{AQN-power}, and Eq.~\ref{numerics1}.    

\exclude{There are  few additional important elements that may increase the likelihood of the axion discovery expressed in terms of parameter $A(t)$ which could be several orders of magnitude larger than its average value for a very short period.}
Furthermore, the estimates in Eq.~\ref{AQN-rate} and Eq.~\ref{AQN-power} explicitly show the presence of enhancement factor $(Lm_a)\gg 1$ which counts the number of layers along the relativistic axion's path, which is a very distinct element in comparison with conventional searches for nonrelativistic axions. This feature implies that if an instrument is asymmetric, e.g. $L_x\gg L_y$ it automatically becomes sensitive to the direction of the axion's momentum, i.e. the instrument potentially becomes directionally sensitive to the AQN-induced axion flux. Therefore, potentially such an instrument allows us to study the DM distribution locally, in our galaxy beyond the conventional SHM paradigm. A hope is that these novel elements in data analysis can overcome the absence of large numerical factor $Q_{\alpha}\sim 10^6$ which is normally present in the cavity resonant type studies.

\subsection{Reasoning behind the expected daily modulation}
\label{sec:modulation}

We now focus on the central aspect of this proposal: the anticipated daily modulation of AQN-induced axions, as described by Eq.~\ref{eq:daily}. This modulation suggests that the count of emitted microwave photons (Eq.\ref{numerics}) or a power excess (Eq.\ref{numerics1}) can accumulate over extended periods, potentially years. To identify a signal, we analyze its presence by fitting the daily modulation pattern given by Eq.~\ref{eq:daily}, summing data across all days of the entire season at a fixed time each day. By examining power excess as a function of time over a 24-hour cycle, any axion-related observables—such as induced currents or voltages in the pickup loop (as reviewed in Sec.\ref{strategy})—should display this same daily modulation, regardless of the specific recording method used, since the expression for $A_{\rm (d)}(t)$ in Eq.~\ref{eq:daily} remains consistent across all observables.

While the emission rate (or power excess) is relatively low, the accumulation of events over several years could yield a statistically significant detection of daily modulations if present in the data. More importantly, this approach can effectively differentiate a genuine signal from spurious signals and noise.

A straightforward method to validate the signal is to observe whether the phase $\phi_0$ in the fitting function of Eq.~\ref{eq:daily} remains constant throughout the year. Instead, within the AQN model, $\phi_0$ should vary slowly with the annual cycle. Specifically, the phase should reach approximately $\phi_0 \simeq \pi$ in opposite seasons (e.g., summer vs. winter), indicating a phase shift where the daily maximum and minimum values invert. This test, which involves comparing frequency patterns in daily and annual modulation data, requires a long observation period (spanning years) but is robust, as systematic errors differ markedly between daily and annual cycles.

Another simple test involves analyzing datasets with ${\cal{B}}=\SI{0}{\tesla}$. The results for ${\cal{B}}=0$ should differ significantly from those obtained with ${\cal{B}} \neq \SI{0}{\tesla}$, helping to distinguish genuine signals from noise.
 
Daily modulations can also be analyzed through Fourier transformation, where a single, well-defined peak at $\Omega_d=2\pi\,\rm day^{-1} $ would indicate a genuine modulation, with smaller, randomly distributed peaks representing noise and other spurious effects. Performing the Fourier transform on ${\cal{B}}=0$ data, in contrast, should yield a more random noise spectrum rather than a clear peak at $\Omega_d=2\pi\,\rm day^{-1} $.
 
An even stronger test for ruling out spurious signals involves using a network of synchronized instruments to detect correlated signals, as discussed in original studies \cite{Budker:2019zka, Liang:2020mnz}.

Various systematic background fluctuations—such as temperature changes, tidal effects, atmospheric phenomena, and human activity—might mimic the daily modulations described here. These spurious signals could be distinguished from genuine axion signals by analyzing the ${\cal{B}}=0$ data and observing phase shifts in $\phi_0$ over different seasons. It would be challenging for spurious signals to replicate all the unique features associated with AQN-induced axions described above.

These considerations suggest that cavity-type experiments—currently the only instruments probing the parameter space for standard QCD axion models with $ 10^{-6} {\rm eV} \lesssim m_a\lesssim 10^{-3} {\rm eV}$ could be used to investigate the daily modulation pattern. These instruments function as non-resonant axion-photon converters within a strong magnetic field.

In principle, any current axion detection instrument could analyze daily modulation following this method, including all previously collected data. This approach allows haloscopes to search for daily modulation by aggregating data (e.g., within a specific month where $\phi_0$ can be assumed constant) for each hour, enabling detection of potential daily modulation patterns.

In this work, we apply this methodology to data collected by the CAST-CAPP detector from September 2019 to June 2021, presenting it as a proof-of-principle demonstration.

\section{Data analysis}
\label{sec:cast_capp_analysis}

CAST-CAPP is an axion haloscope that operated from September 2019 to June 2021 at CERN in Geneva, Switzerland. It consists of four tuneable $23 \times 25 \times 390$\si{\mm} rectangular cavities, each with a volume of $V = \SI{224}{\centi\meter\cubed}$ which were placed inside one of the two bores of CAST's superconducting dipole magnet of \SI{8.8}{\tesla}. CAST-CAPP introduced two novel mechanisms to increase its sensitivity to DM axions but also to transient events:
\begin{itemize}
    \item The phase-matching technique combines  the power outputs from the four identical cavities coherently to increase the Signal-to-noise ratio (SNR) linearly with the number of cavities.
    \item The state-of-the-art fast-tuning mechanism with a speed of \SI[per-mode = symbol]{10}{\MHz\per\minute} and a resolution of \SI{100}{\Hz}, together with wideband electronics allow for a search for transient events such as axion streams and axion miniclusters.
\end{itemize}

CAST-CAPP acquired 172 days of data using both single and phase-matched cavities in data-taking conditions. The total covered frequency range extended from \SIrange{4.77}{5.43}{\GHz} covering a parameter phase space of $\sim$$\SI{660}{\MHz}$ corresponding to axion masses between \SIrange{19.74}{22.47}{\micro\eV}. This allowed CAST-CAPP to exclude axion-photon couplings for virialized galactic axions down to $ g_{a\gamma\gamma} = 8 \times 10^{-14}\,\si{\GeV\tothe{-1}} $ at $90\%$ confidence level \cite{castcapp:2022}.

The design and the data-taking scheme of the CAST-CAPP experiment were not originally optimized for a search of daily modulated signals. However, the large acquired bandwidth of \SI{5}{\MHz} compared to the resonance width of about \SI{200}{\kHz}, together with the wide frequency range of $\SI{660}{\MHz}$, allows a first proof of principle demonstration of such a novel analysis procedure.

Finally, it is mentioned that for the probed axion mass range \SIrange{19.74}{22.47}{\micro\eV}, the corresponding Compton wavelength $\lambda_C = {h}/{m_a c}$ ranges from approximately \SIrange{6.3}{5.5}{\cm}. Given the length of each single rectangular cavity $L = \SI{39}{\cm}$, the condition $m_a L$ yields:
\begin{equation}
\label{eq:capp_mL_condition}
m_a L \approx \frac{1}{\lambda_C} \cdot L \approx 6.2 \text{ to } 7.1.
\end{equation}
Thus, the condition $m_a L \gg 1$  being used in our estimates is marginally satisfied for the entire frequency range of CAST-CAPP, indicating that the cavity length is sufficiently long to justify the requirement for relativistic axion detection.

\subsection{Simulation}
\label{sec:simulation}

The AQN signal search is simulated using artificially modulated signals incorporated into simulated spectra. These spectra are composed of randomly generated noise superimposed onto authentic spectral baselines acquired from CAST-CAPP. The CAST-CAPP data are recorded in \SI{1}{\minute} bins which are then converted into frequency-domain using a fast Fourier transformation. Each spectra carry a unique shape originating from the $TE_{101}$ resonance mode of the cavities and the amplifier gain profile. To preserve this unique shape in the simulated spectra, we produce spectral baselines using a low-pass filter. Each simulated spectrum is comprised of artificial Nyquist noise multiplied by a randomly selected spectral baseline. The artificial noise has a Gaussian distribution of $\mu=1$ and $\sigma=1/\sqrt{\text{resolution bandwidth} \cdot \text{spectrum time length}} = 0.018$. Then, the artificial daily modulated signal is simulated using a sine function of tuned amplitude added on top of the artificial noise spectrum.

\begin{figure}[htb!]
    \centering
    \includegraphics[width=\linewidth]{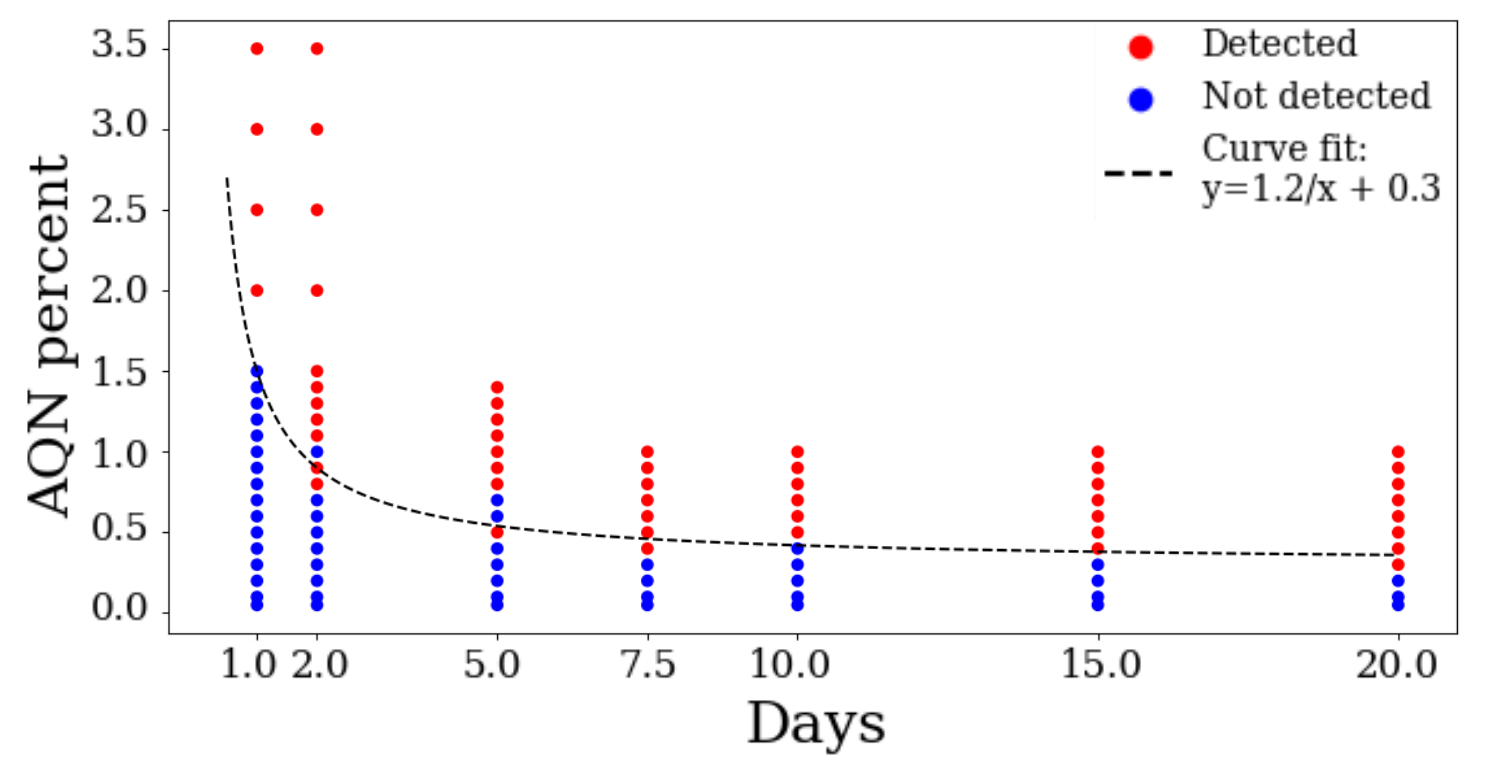}
    \caption{Simulation of AQN signals with different power percentages as a function of data taking time. The red points indicate discoveries while the blue ones denote insignificant signals. A decision boundary fit is denoted by the black dashed line.}
    \label{fig:simulation_aqn_power}
\end{figure}

\begin{figure}[htb!]
    \centering
    \includegraphics[width=\linewidth]{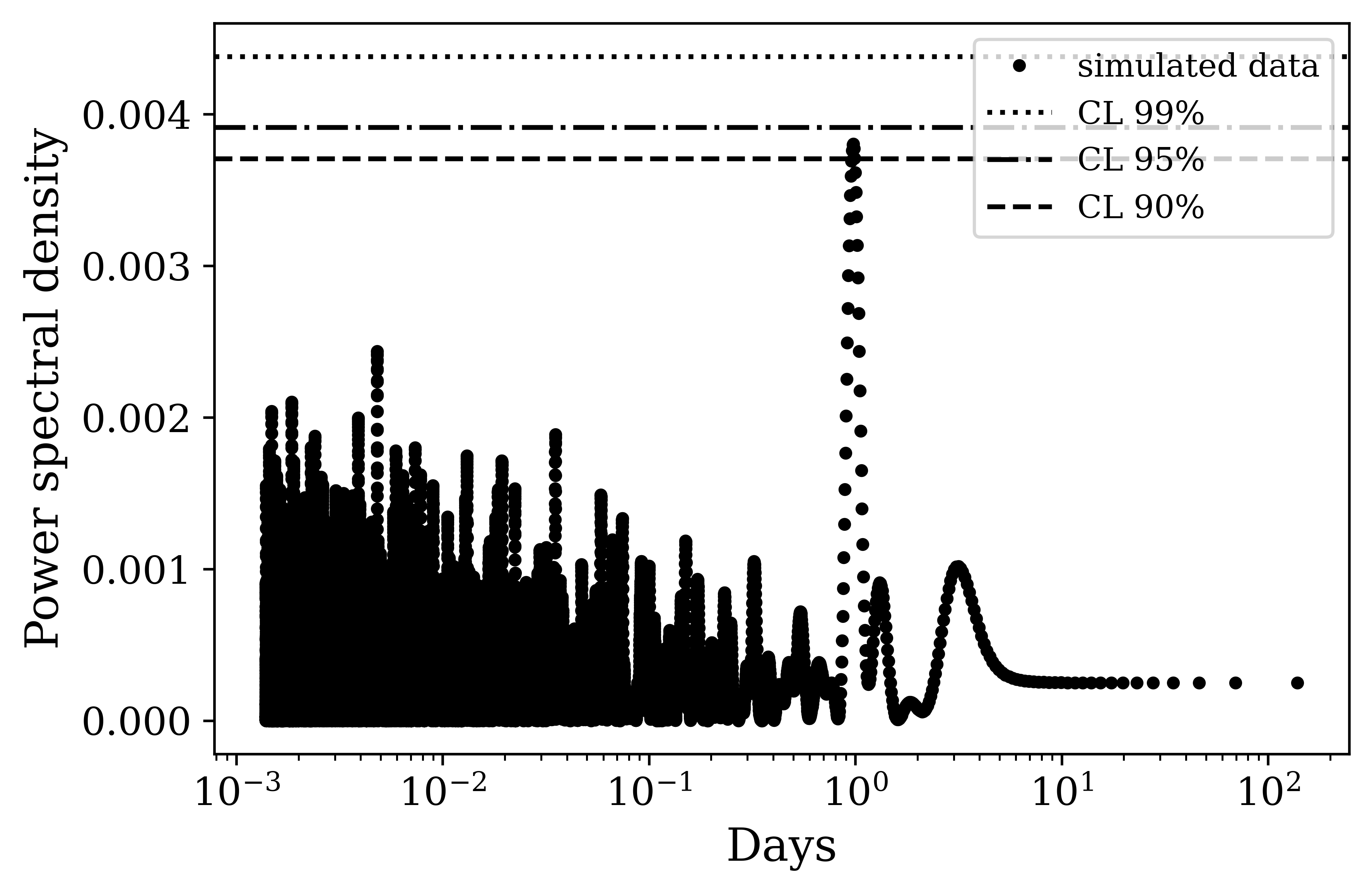}
    \caption{Periodogram created using the Lomb Scargle method with 5 days of simulated data taking time (5 days / 1 min = 7200 simulated spectra). The observed signal is above 90\% CL.}
    \label{fig:simulation_periodogram}
\end{figure}

We performed multiple simulations with the data-taking time ranging from 1 day to 20 days at a single frequency and an AQN signal amplitude ranging from $0.01\%$ to $3.5\%$ of the mean spectral power. Figure \ref{fig:simulation_aqn_power} shows the combinations of simulation parameters i.e. number of data taking days and AQN amplitude percentage with respect to the mean spectral power on a two-dimensional (2D) space, while the color denotes if that specific combination yielded a significant signal that can be detected on a Lomb Scargle periodogram. The confidence level for the significance is set to be $90\%$. In Fig.~\ref{fig:simulation_periodogram}, a periodogram for a significant AQN signal is shown for 5 days of data taking. In this case, the power of the AQN signal is adjusted to be $0.8\%$ of the mean spectrum power.

\subsection{Data selection}
\label{sec:data_selection}

During its $\sim$$2$ years of operation, CAST-CAPP recorded  64.3 days of data using phase-matched cavities and 107.5 days of data using a single cavity. At the same time, a total of 16.4 days of data were recorded with ${\cal{B}}=\SI{0}{\tesla}$. The total frequency range of CAST-CAPP is \SI{644}{\MHz} between \SIrange{4.79}{5.43}{\MHz}.

The strategy for quick identification of parasitic environmental background electromagnetic signals involved the usage of an external omnidirectional antenna that was placed outside the CAST magnet and was connected to a spectrum analyzer recording data simultaneously with the cavities. This Electromagnetic Interference / Electromagnetic Compatibility (EMI/EMC) identification system was introduced in November 2020 and collected $89.2$ d of data.

For the search for a daily modulation with the CAST-CAPP data, only the data acquired from single cavities are used. The reason is that for the phase-matching procedure, a series of programmable phase-independent attenuators were used to adjust the amplitude of the resonance of the individual cavities with \SI{0.25}{\dB} to match the resonance peaks of the phase-matched cavities. This introduced additional attenuation although it is not expected to follow a daily rhythm, it could introduce unnecessary artifacts in the whole procedure.

Additionally, the data that were recorded during the usage of the fast-frequency scanning search approach are omitted from the analysis due to the inherent frequency dependency of the amplitude of the resonance peak of the $TE_{101}$ mode. 

Therefore, as a first approximation only data taken from single cavities at fixed frequencies are used. Because of the high mechanical $Q$-factor of the sapphire strips of the tuning mechanisms, which were accidentally acting as a mechanical tuning fork, a frequency modulation of the electromagnetic mode of interest was observed in certain frequencies. The data that were dominated by such mechanical vibrations as well as other undesired non-systematic effects or possible EMI/EMC interferences were excluded from further consideration using a series of selection criteria. 

\subsection{Data treatment}
\label{sec:data_treatment}

Each CAST-CAPP spectrum that fulfills the aforementioned cuts is averaged within itself to get the mean spectral power, then a timestamp is attached to it. This way, a 2D array of mean spectral powers vs corresponding acquisition timestamps is created. This array is then rebinned hourly to have mean spectral powers for each hour of data acquisition.

\begin{figure}[htb!]
    \centering
    \includegraphics[width=\linewidth]{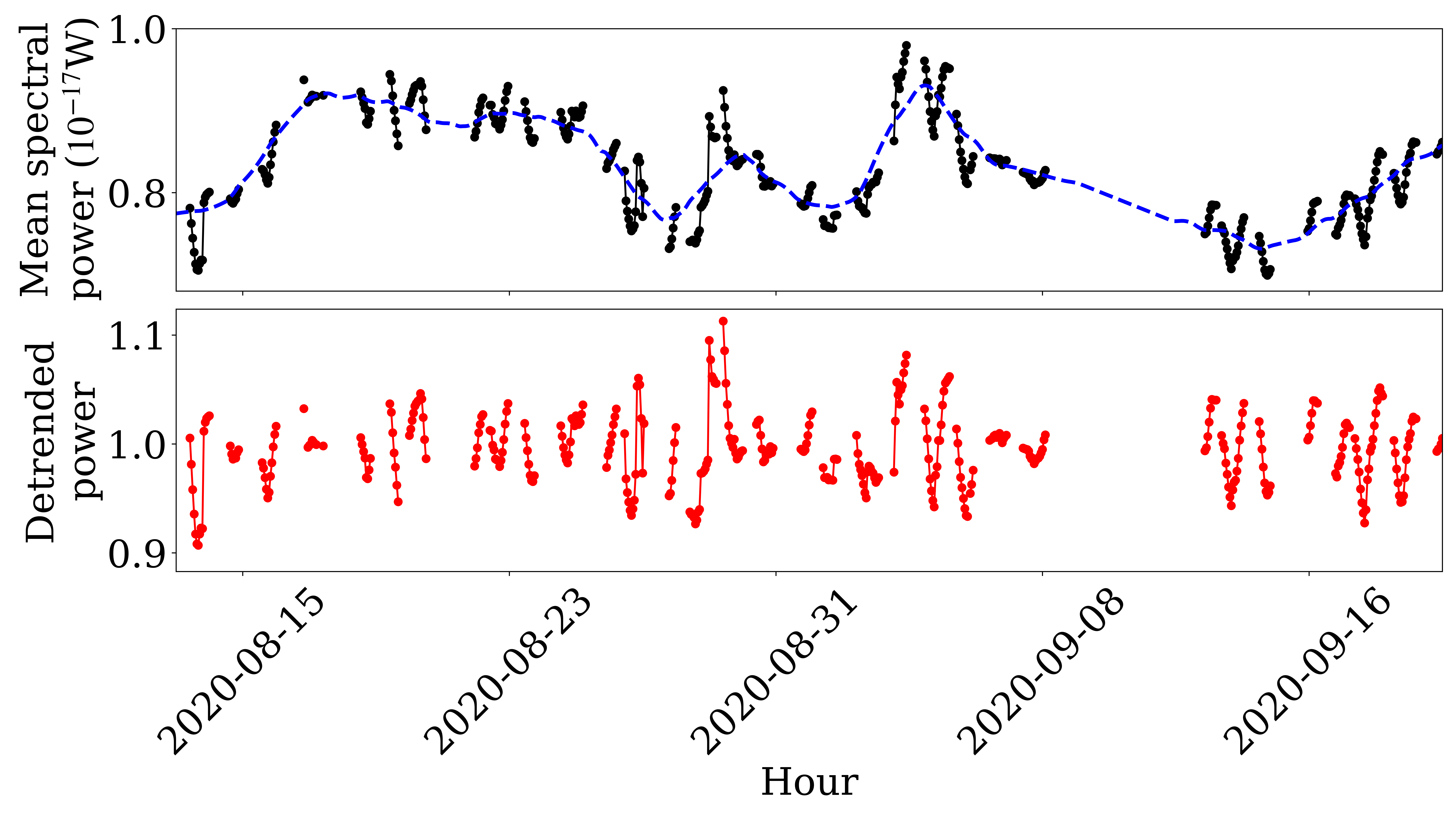}
    \caption{Mean hourly spectral power (black) of cavity 3 between August 13 and September 20. The blue dashed curve denotes the trend of the data created by the low-pass filter. The detrended power (red) is calculated by dividing data (black) by its trend (blue).}
    \label{fig:detrend}
\end{figure}

The baseline on which the daily modulation in the CAST-CAPP data are investigated is expected to fluctuate due to weekly environmental temperature changes as well as amplifier gain drifts. To remove this effect before the daily modulation analysis, the data are detrended. For this purpose, a Savitzky Golay filter is used with a window length of two days and a polynomial order zero. Figure \ref{fig:detrend} shows an example of the mean spectral power time-series belonging to cavity 3 between August 13, 2020 and September 20, 2020, with ${\cal{B}}=\SI{8.8}{\tesla}$ before and after the detrending procedure. The parameters of the low-pass filter are selected such that the long-range fluctuations are removed while the investigated daily variations are conserved. After detrending, the mean spectral variations are around a baseline that equals 1.

\subsection{Systematics}
\label{sec:systematics}

CAST-CAPP's experimental setup consists of a three-stage signal amplification using High-electron-mobility transistor (HEMT) amplifiers. The first amplification for each cavity was performed close to the critically coupled signal output port at the \SI{4}{\kelvin} stage using a HEMT low-noise amplifier (LNA) from Low Noise Factory (LNF-LNC4\_8D) providing about \SI{39}{\dB} amplification with \SI{2}{\kelvin} noise for the frequency of CAST-CAPP at data-taking conditions. Then, an external amplification of about \SI{22}{\dB} for each of the four cavities is performed using four room-temperature ZX60-83LN-S+ LNAs. Finally, a third-stage amplification is performed at the output signal after the signal combination using a Miteq AFD3-0208-40-ST LNA providing about \SI{30}{\dB} gain.

\begin{figure*}[t]
    \centering
    \includegraphics[width=0.9\linewidth]{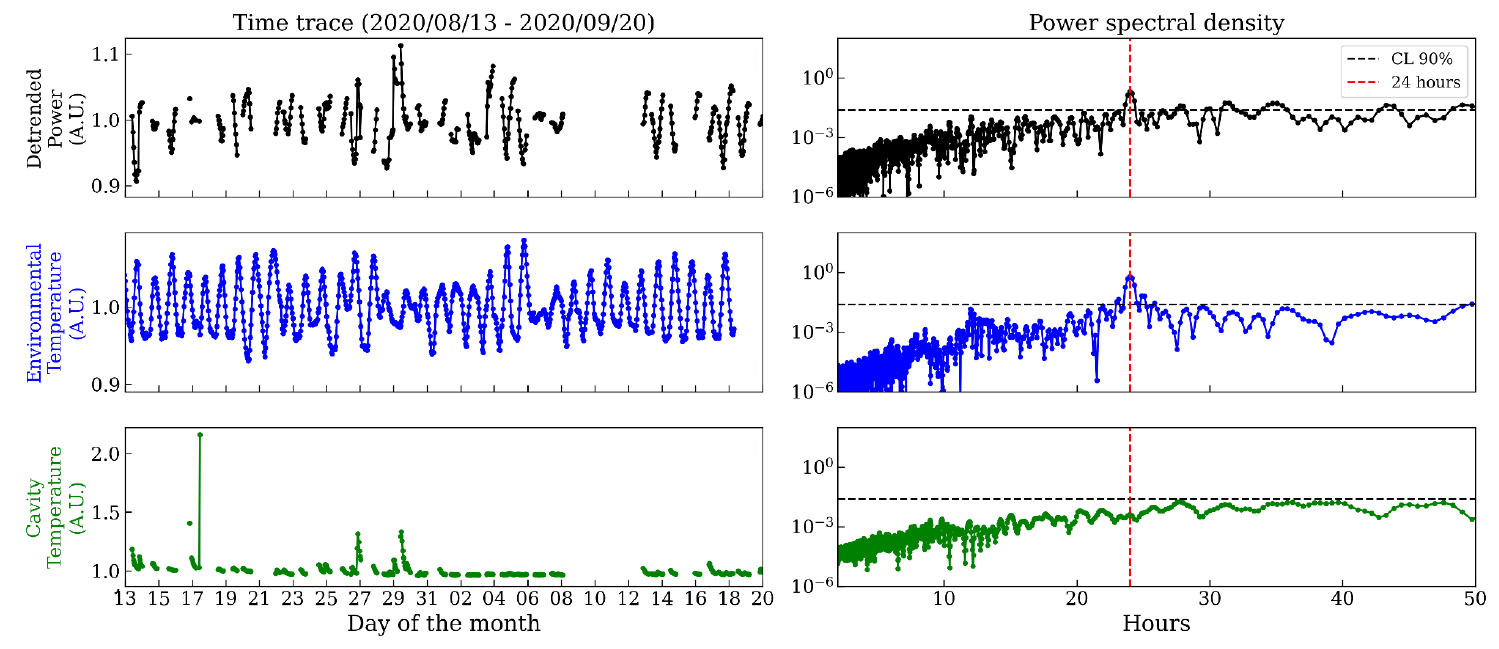}
    \caption{Time series (left) and Lomb Scargle periodograms (right) for data taken between August 13, 2020 and September 20, 2020, with cavity 3 and $B=\SI{8.8}{\tesla}$. The results for the ambient environmental temperature as well as the cavity temperature are also shown.}
    \label{fig:periodogram_bon}
\end{figure*}

Two main parameters that could influence the performance of these amplifiers are the provided bias voltage as well as the temperature. Regarding the bias voltage of the room-temperature amplifiers, this was provided using a Hameg HM7044 power supply. The accuracy of this power supply is on the level of \SI{10}{\milli\volt} which translates into a tiny gain variation of about \SI{0.007}{\dB} or $\sim$$0.2\%$ in terms of power for the second- and third-stage amplification. For the first-stage amplification, a LNF-PS\_3 was used for the biasing (\SI{0.25}{\volt} and \SI{7}{\milli\ampere}) for each one of the four LNAs corresponding to the four cavities. The biasing conditions of the LNAs were optimized for minimal heat dissipation and minimum noise figure. A current variation of about \SI{0.02}{\milli\ampere} was observed which would translate into a variation of the gain on the level of \SI{0.02}{\dB} or $\sim 0.5\%$ in terms of power.

The temperature of the first stage LNA was monitored through Cernox temperature sensors to be varied between \SIrange{6}{10}{\kelvin}. For the specific bias of about \SI{0.25}{\volt} and \SI{7}{\milli\ampere} at a frequency of $\sim$$\SI{5}{\GHz}$ the gain of LNF-LNC4\_8D has a linear dependency on temperature, increasing the gain of about \SI{0.25}{\dB} and an increase of their noise temperature \SI{0.42}{\kelvin} for the aforementioned temperature range \cite{lnf_private}. This translates into a $\sim 5.9\%$ variation of the gain in terms of power.

The temperature and humidity of the experimental area of CAST were monitored and kept stable via a central ventilation and temperature control system. However, small changes of a few degrees were observed. More specifically, the temperature variation observed in the room-temperature electronics of CAST-CAPP during the whole data-acquisition period was measured using PT100 Resistance Temperature Detector (RTD) temperature sensors to be between \SIrange{19}{26}{\degreeCelsius}. These small temperature changes result in a linear gain dependency of the room-temperature amplifiers. For the second stage amplification a temperature change between \SIrange{19}{26}{\degreeCelsius} results in a decrease of the gain of about \SI{0.078}{\dB} or about $1.8\%$ decrease in power \cite{minicircuits_private}. Then, the third stage amplifier is expected also to have a decrease of its gain with increasing temperature of about \SI[per-mode = symbol]{0.02}{\dB\per\degreeCelsius} \cite{miteq_private}. Therefore, for \SIrange{19}{26}{\degreeCelsius}, the third-stage room-temperature amplifier would exhibit a decrease of $\sim \SI{0.14}{\dB}$ or about $3.3\%$ decrease in terms of power.

It is noted that the gain stability of the amplifiers over time is expected to have a much smaller variation than the changes in temperature, voltage, or other ambient conditions. An overview of the aforementioned systematics and their effect on the gain of the various amplifiers of CAST-CAPP is shown in Table \ref{tab:systematics}.

\begin{table}[htbp]
  \centering
  \begin{tabular}{c|c|c|c}
        &  \textbf{First stage}  &  \textbf{Second stage} &  \textbf{Third stage} \\
    \hline
    Temperature  & $+5.9\%$             & $-1.8\%$            & $-3.3\%$  \\
    Bias      & $-0.5\%$              & $+0.2\%$            & $+0.2\%$  \\
  \end{tabular}
  \caption{Systematic effects on the variation of power from the various amplification stages of CAST-CAPP. The $\pm$ signs indicate a positive or a negative correlation with the corresponding parameter.}
  \label{tab:systematics}
\end{table}

Finally, the magnetic field of CAST's dipole magnet was measured to have a stability of \SI{0.1}{\milli\tesla} at the level of \SI{8.8}{\tesla}, translating into an insignificant $0.001\%$ variation.

\begin{figure*}[t]
    \centering
    \includegraphics[width=0.9\linewidth]{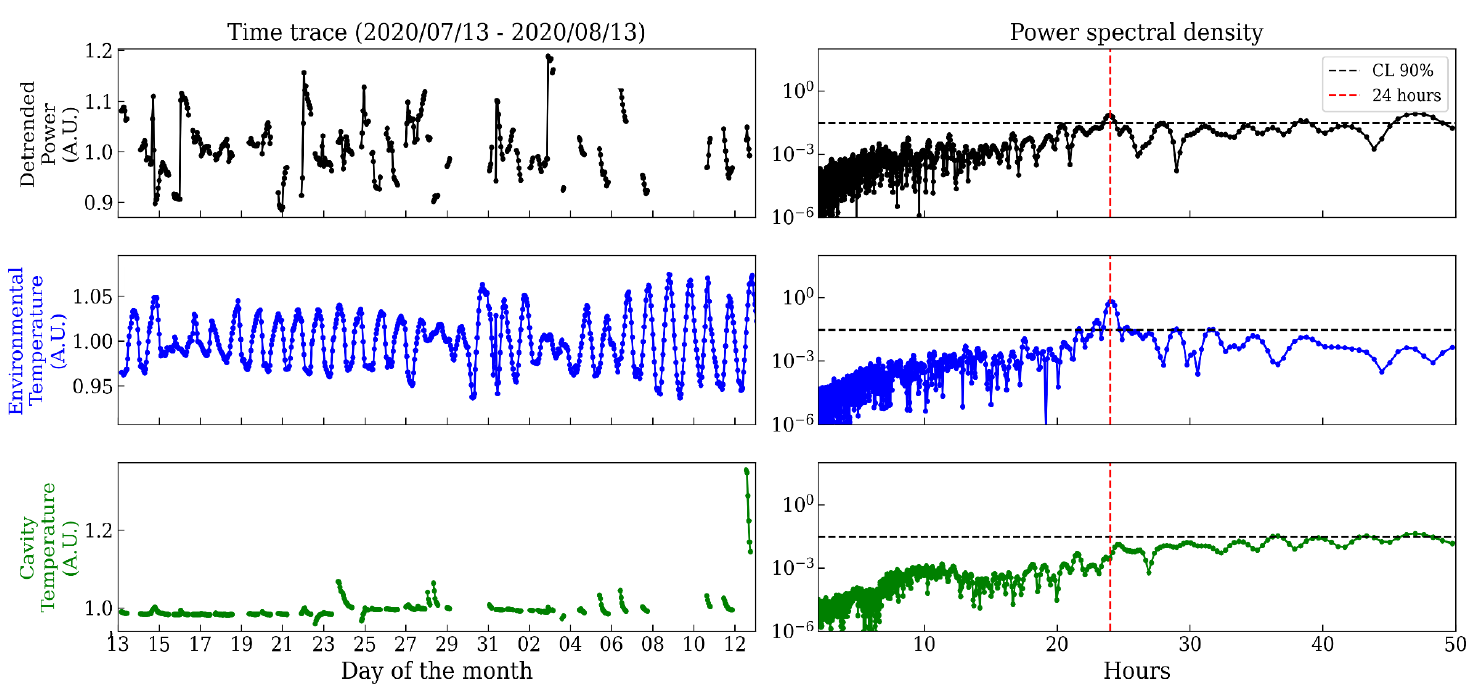}
    \caption{Time series (left) and Lomb Scargle periodograms (right) for data taken between July 13, 2020 and August 13, 2020 with cavity 2 and $B=\SI{0}{\tesla}$. The results for the ambient environmental temperature as well as the cavity temperature are also shown.}
    \label{fig:periodogram_boff}
\end{figure*}

\subsection{Results}

\subsubsection{Lomb Scargle periodogram}

The first step for a search for daily periodicities is performed using a Lomb-Scargle periodogram in the detrended data. This way periodicities around \SI{24}{\hour} can be investigated. Figure \ref{fig:periodogram_bon} shows the time series (top left) as well as the periodogram (top right) for ${\cal{B}}=\SI{8.8}{\tesla}$, where the time period from August 13, 2020 to September 20, 2020 is selected for single cavity 3. At the same time, using the same period, the environmental temperature from the PT100 RTD temperature sensors (Fig.~\ref{fig:periodogram_bon} middle row) as well as the cavity temperature from the Cernox sensors (Fig.~\ref{fig:periodogram_bon} bottom row) are shown. Table \ref{tab:data_stats} gives a detailed account of the data statistics. The observed gaps in the time series for both the cavity power and the cavity temperature data are due to CAST-CAPP's data acquisition scheme not being optimized for a daily modulation search, i.e. CAST-CAPP acquired single-cavity data for a limited and noncontinuous amount of time. As the time series data were detrended and normalized, the y axes are given in arbitrary units. In the periodograms (Fig.~\ref{fig:periodogram_bon} right) that show the power spectral densities of the corresponding time series, the red dashed vertical line denotes the daily periodicity whereas the black dashed horizontal line shows the threshold for a significant signal with a confidence level of $90\%$.

\begin{table*}[htbp]
  \centering
  \footnotesize
  \begin{tabular}{c|c|c}
                                 & \textbf{${\cal{B}}=\SI{8.8}{\tesla}$ (Fig.~\ref{fig:periodogram_bon})}  & \textbf{${\cal{B}}=\SI{0}{\tesla}$ (Fig.~\ref{fig:periodogram_boff})} \\
    \hline
    Interval                     & August 13 - September 20                & July 13 - August 13 \\
    Data amount (1-min files)    & 14400                          & 11500  \\
    Total hours                  & 240                            & 192 \\
    Hours per day                & 8.0                            & 5.2 \\
    Frequency range [\si{\GHz}]  & 5.20 - 5.33                    & 5.32 - 5.41 \\
  \end{tabular}
  \caption{Data statistics for Figs.~\ref{fig:periodogram_bon} and \ref{fig:periodogram_boff}.}
  \label{tab:data_stats}
\end{table*}

As expected the environmental temperature follows a daily rhythm while the cavity, and therefore the cryogenic LNA temperature, does not. Interestingly, also the detrended cavity data for ${\cal{B}}=\SI{8.8}{\tesla}$ show a \SI{24}{\hour} periodicity above $90\%$ confidence level. However, the ambient temperature change can be the leading cause behind the observed daily variation due to its direct effect on the receiver chain, as mentioned in the previous section. This is due to the room-temperature setup not being sufficiently isolated from temperature changes.

For comparison reasons, also, ${\cal{B}}=\SI{0}{\tesla}$ data are analyzed. For cavity 2, these are shown in Fig.~\ref{fig:periodogram_boff} with the selected period ranging from July 13, 2020 to August 13, 2020. For practical reasons, this period could not be the same as the period for the ${\cal{B}}=\SI{8.8}{\tesla}$, and therefore, a perfect comparison between the two cannot be made. The time difference between the two is about 1 month. As can be seen, these data with ${\cal{B}}=\SI{0}{\tesla}$ also indicate a daily modulation being less pronounced than the ${\cal{B}}=\SI{8.8}{\tesla}$ data, which may be attributed to the ${\cal{B}}=\SI{0}{\tesla}$ data being more sparse compared to ${\cal{B}}=\SI{8.8}{\tesla}$ data. Therefore, we conclude that the appearance of similar daily modulation in the detrended power spectra of Figs.~\ref{fig:periodogram_bon} (${\cal{B}}=\SI{8.8}{\tesla}$) and \ref{fig:periodogram_boff} (${\cal{B}}=\SI{0}{\tesla}$), does not allow to exclude systematics for the hint in Fig.~\ref{fig:periodogram_bon} but strengthens the assumption of a temperature-dependent amplifier gain effect. Of course, there is always the possible dark photon contribution with no magnetic field dependence.

\subsubsection{Hourly distributions}

\begin{figure*}[htb!]
    \centering
    \includegraphics[width=0.8\linewidth]{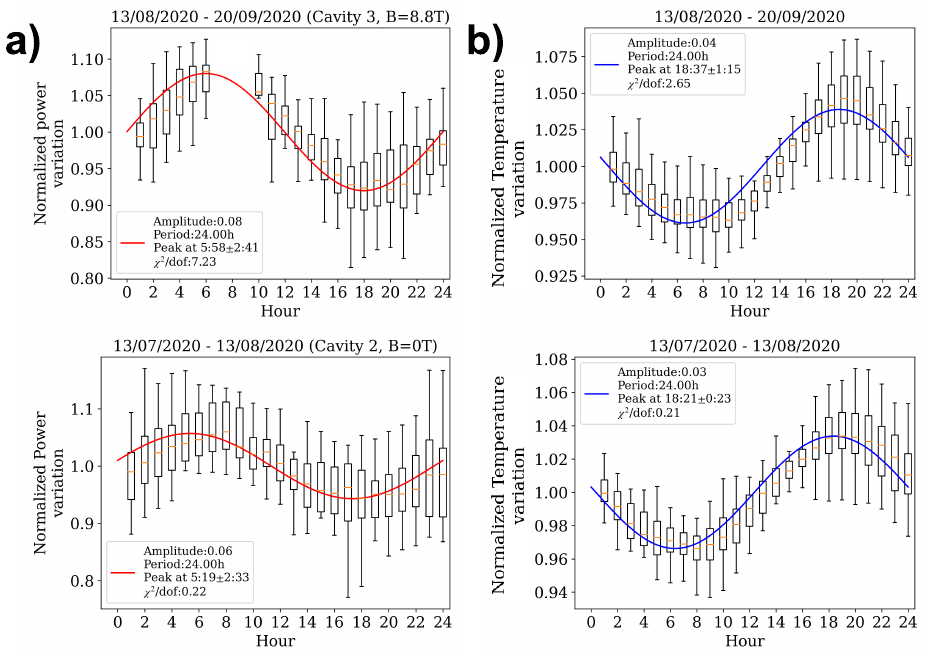}
    \caption{Box plots showing the hourly distributions of the normalized power variation (left) and normalized environmental temperature variation (right) for ${\cal{B}}=\SI{8.8}{\tesla}$ (top) and ${\cal{B}}=\SI{0}{\tesla}$ (bottom). A sine function with a daily periodicity is fitted to the data. The gap in the upper left plot is due to the CAST-CAPP data acquisition schedule in that time interval of data taken.}
    \label{fig:box_plots}
\end{figure*}

Next, the hourly distributions of both ${\cal{B}}=\SI{8.8}{\tesla}$ and ${\cal{B}}=\SI{0}{\tesla}$ data are shown in Fig.~\ref{fig:box_plots} which could provide more information on the observed daily modulation. These distributions are derived for the same time intervals as in Fig.~\ref{fig:periodogram_bon} and Fig.~\ref{fig:periodogram_boff}, and are shown in the form of box plots in Fig.~\ref{fig:box_plots}a, top and bottom, respectively. A sine function is then fitted to the data. The maxima of the fitted sinus shapes are found around 05:58 and 05:19 local hour, for ${\cal{B}}=\SI{8.8}{\tesla}$ and ${\cal{B}}=\SI{0}{\tesla}$ respectively.

Even assuming the sine-wave fit of $8.1\% \pm 0.6\%$ and $6.0\% \pm 0.5\%$ for the mean daily distributions of ${\cal{B}}=\SI{8.8}{\tesla}$ and ${\cal{B}}=\SI{0}{\tesla}$ respectively, the CAST-CAPP data do not allow for a significant conclusion in favor of the AQN-related axion signal.

The hourly temperature distributions corresponding to Fig.~\ref{fig:box_plots}a are shown in Fig.~\ref{fig:box_plots}b where we observe that the temperature is maximum at around 18:30 Central European Time (CET). This corresponds approximately to half a day phase shift between the mean spectral power and the ambient temperature variations shown in Fig.~\ref{fig:periodogram_bon} and can be explained by the argumentation in Sec.~\ref{sec:systematics}. 

Table \ref{tab:data_stats_2} gives an overview of the sine fit characteristics for the box plots shown in Fig.~\ref{fig:box_plots}. The sine fits, as shown by chi-squared tests, are significantly more suitable to the data compared to linear fits, confirming the daily periodicity observed. However, it is apparent that the daily change in the environmental temperature of the CAST hall does not allow for a positive conclusion.

\begin{table}[htb!]
  \centering
  \begin{tabular}{c|l|c|c}
    \multicolumn{2}{c|}{ }  &  \textbf{Cavity} &  \textbf{Environmental}     \\
    \multicolumn{2}{c|}{ }  &  \textbf{power}  &  \textbf{temperature} \\
    \hline
           & Fluctuation  & 8.1\% $\pm$ 0.6\%    & 3.7\% $\pm$ 0.3\%         \\
    {\cal{B}}=8.8T & max hour    & 5:58$\pm$3:05 CET  & 18:37$\pm$1:15 CET       \\
           & $\chi^2$/dof & 7.2(sine), 19.2(line)    & 2.6         \\
    
    \hline
           & Fluctuation  & 6.0\% $\pm$ 0.5\%    & 3.4\% $\pm$ 0.3\%         \\
    {\cal{B}}=0T   & max hour    & 5:19$\pm$2:33 CET   & 18:21$\pm$0:23 CET       \\
           & $\chi^2$/dof & 0.2(sine), 8.3(line)   & 0.21        \\
    \hline
  \end{tabular}
  \caption{Data statistics for Fig.~\ref{fig:box_plots}.}
  \label{tab:data_stats_2}
\end{table}

\begin{figure}[htb!]
    \centering
    \includegraphics[width=0.9\linewidth]{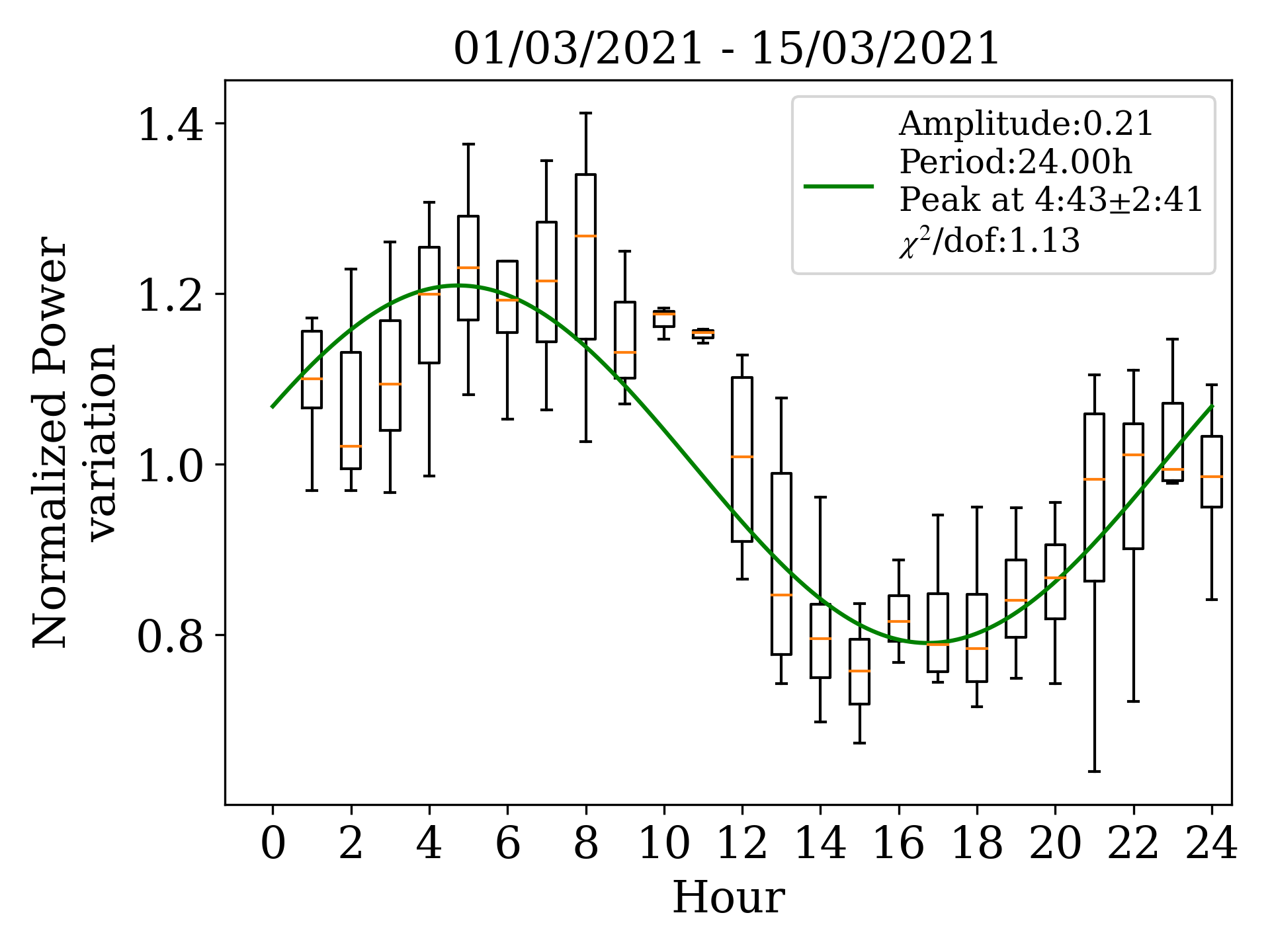}
    \caption{Box plot of the external antenna spectral power.}
    \label{fig:box_plots_antenna}
\end{figure}

Finally, as described in \cite{castcapp:2022}, an external quasiomnidirectional antenna was placed next to the CAST magnet and operated simultaneously with the Spectrum Analyzer (SA) connected to the cavities and at the same frequency band. This was used as a fast veto procedure where a candidate axion signal could be easily discarded if it appeared both in the cavities and in the witness channel at the same frequency. Therefore as an additional comparison, the hourly distribution of the external antenna data is shown in Fig.~\ref{fig:box_plots_antenna}. As expected due to the lack of any temperature isolation, the mean daily spectral variation observed is about $21\%$, larger compared to both ${\cal{B}}=\SI{8.8}{\tesla}$ and ${\cal{B}}=\SI{0}{\tesla}$ data acquired by a receiver chain that is partially temperature isolated.

\subsubsection{Seasonal phase shift}
\label{seasonal_phase_shift}

Because of the limited amount of data taken with CAST-CAPP, the expected seasonal phase shift (see Sec.~\ref{sec:modulation}) has been searched for, but could not be identified. The periods August 13, 2020, to September 20, 2020, and November 18, 2020, to December 02, 2020 have been analyzed, where the expected phase shift is only of the order of $0.3 \pi$ or 3 hours and 40 minutes, however, due to the limited statistics, this number could not be determined.

\section{Conclusion and future studies}
\label{conclusion} 

In this work, as a proof of principle study, we have reanalyzed the CAST-CAPP data already published in \cite{castcapp:2022}, focusing on the daily modulation of a potential axion signal due to DM axions or from AQNs as it has been presented in Sec.~\ref{signals}. We have observed daily variations, but at the same time also for ${\cal{B}}=\SI{0}{\tesla}$ (see Fig.~\ref{fig:box_plots} bottom) and with the external antenna (see Fig.~\ref{fig:box_plots_antenna}). Since a similar dependence is expected for the electronic gain of the amplifier chain of the Data Acquisition (DAQ) system, we could not conclusively determine an axion-related signal (whatever its origin: from streams or AQNs), or a signal from the hidden sector. Additional tests to discriminate a true genuine signal from the noise or spurious signals cannot be carried out, as the CAST-CAPP detector is no longer operational. Although the data in Fig.~\ref{fig:box_plots} do show some daily oscillation, we conclude that this could be due to temperature effects. However, we can formulate some lessons to be learned, which can be useful for future studies and different instruments, yet to be built: 

\begin{enumerate}

\item The unavoidable fluctuations of the electronic gain of the amplifier chain, in radiometers, have led to the invention of the Dicke switch to eliminate such gain fluctuations of the entire chain, and not only the front-end like for the noise figure. The inherent switching noise added by this type of radiometer can then be mitigated by the use of the correlation-type radiometer \cite{Tanner-2022-applications, Menon-2019-generalized}. Virtually every modern microwave remote-sensing satellite is based on this correlation technique, and therefore axion DM experiments can also benefit from this approach and replace the total power radiometer that is currently widely used. This approach can allow for a conclusive search for daily modulations in the future.

\item The expected AQN-related phase shift during different seasons can also serve as an additional check to distinguish a genuine AQN signal from purely environmental or terrestrial influences.

\item An additional unequivocal verification of the existence of an external AQN-related modulation can be derived from the search for sidereal periodicities. These periodicities are anchored to a fixed referenced frame relative to remote stars rather than the Sun thus indicating an exosolar modulation source. The sidereal day of 23 h 56 min and the Moon's sidereal month of 27.32 days are characteristic examples of such periodicities. With high-quality data with enough statistics, this sidereal modulation can become statistically resolvable, thus linking the potential signal to cosmic rather than local environmental sources. 

\item To determine the mass and other parameters of the AQN-originated axion a truly broadband instrument should be used. This should be contrasted with the canonical resonant type of experiments when the positive observation of a signal in the form of a narrow peak at $\nu_0$ unambiguously predicts the axion mass $m_a=\nu_0*(2\pi)$ irrespective of any normalization factors such as the DM density. However, in the case of the nonresonant studies as advocated in this work, the axion mass cannot be extracted with sufficiently high precision as $m_a$ enters only as a normalization factor (\ref{numerics1}) and is highly sensitive to many uncertainties in the estimates including the general normalization factor such as the local DM density and its possible deviations from the SHM as mentioned above. The local DM density can also vary due to possible sidereal effects as mentioned above.

\item For cavity-type experiments, the sensitivity to higher-frequency photons in broadband detection is limited by the specific coupling and the characteristics of the amplifier chain, whose gain and noise figure are typically constrained to several GHz. Additionally, the bandwidth of the readout system acts as a limiting factor, effectively functioning as a bandpass filter. Future instruments must address these limitations by designing amplifier chains and readout electronics with significantly broader frequency coverage to detect signals from AQN-produced axions.

\item Finally, the best option to discriminate the true genuine signal from the noise or spurious signals is to analyze the correlations in two or more different instruments representing a synchronized network, as suggested in \cite{Budker:2019zka, Liang:2020mnz}. The basic idea is to search for some transient events that occur at (almost) the same instant in the different instruments of a network of axion detectors. The typical rate for such correlated events depends on the strength of the spikes as discussed in \cite{Budker:2019zka, Liang:2020mnz}. 

\end{enumerate}

If future experiments conclude that the AQN-induced axions are responsible for daily modulation, it would have profound consequences. It would imply that two long-standing puzzles in cosmology, the nature of the DM and the matter-antimatter asymmetry of our Universe are simultaneously resolved. These two puzzles are intimately linked within the AQN framework and an observed signal cannot be explained by any other model since such a signal could be only generated by the relativistic AQN-induced axions (which are mostly localized on the Earth's surface in contrast with CaB\footnote{As we mentioned in Sec.~\ref{sect:introduction} the local density of the AQN-induced axions on the Earth's surface should be at least 4 orders of magnitude higher than the density of the cosmologically produced  CaB axions considered in Ref.~\cite{Dror:2021nyr}; see the paragraph above Eq.~\ref{flux} for an estimate. This is because the CMB photon density strongly constrains CaB density. The spectral features of the AQN-induced axions are also vastly different from the CaB axions.}).

\section*{Acknowledgments}
  
This work has been supported by the Greek General Secretariat for Research Innovation (GSRI), the National Science and Engineering Research Council of Canada, the Institute for Basic Science (IBS) under Project No. IBS-R017-D1 of the Republic of Korea, the Spanish Agencia Estatal de Investigacion (AEI) and Fondo Europeo de Desarrollo Regional (FEDER) under projects No. FPA-2016-76978-C3-2-P and No. PID2019-108122GB-C33, by the CERN Doctoral Studentship programme, the European Research Council (ERC), the DFG Research Training Group Programme 2044 “Mass and Symmetry after the Discovery of the Higgs Particle at LHC”, MSE (Croatia) and Natural Sciences and Engineering Research Council of Canada. We acknowledge support through the ERC under Grants No. ERC2018-StG-802836 (AxScale project) and No. ERC-2017-AdG-788781 (IAXO+ project). Part of this work was performed under the auspices of the U.S. Department of Energy by Lawrence Livermore National Laboratory under Contract No. DE-AC52-07NA27344. This article is based upon work from COST Action COSMIC WISPers CA21106, supported by COST (European Cooperation in Science and Technology).

\bibliography{main}

\end{document}